\documentclass[12pt,prd,onecolumn,showpacs,amsmath,amssymb,aps,floats,floatfix,nofootinbib]{revtex4-1}

\usepackage[colorlinks=true,urlcolor=blue,anchorcolor=blue,citecolor=blue,filecolor=blue,linkcolor=blue,menucolor=blue,linktocpage=true]{hyperref} % should be commented out if the tex file will be compiled with latex in arXiv!!! (pdflatex is fine)

\usepackage[inline]{enumitem}
\usepackage{dcolumn}
\usepackage{bm}
\usepackage{amsmath}
\usepackage{amsfonts}
\usepackage{amssymb}
\usepackage{color}
\usepackage{float}
\usepackage{latexsym}
\usepackage{slashed} % slash mark
\usepackage{pstricks}
\usepackage{indentfirst}
\usepackage{mathrsfs}
\usepackage{multirow}
\usepackage{epsfig,psfrag}
\usepackage{subfigure}
\usepackage{mathtools}
\usepackage{setspace} % spacing
\usepackage[utf8]{inputenc} % accept utf-8 input encoding
\usepackage[scientific-notation=true]{siunitx} % comprehensive units

\graphicspath{{fig/}}

\allowdisplaybreaks % allow eqnarray breaks
\begin{document}

\title{Dark mater interactions from an extra $\mathrm{U}(1)$ gauge symmetry with kinetic mixing and Higgs charge}
\author{Lianyou SHAN}\email{shanly@ihep.ac.cn}
\affiliation{University of Chinese Academy of Sciences (UCAS),  Beijing, China}
\affiliation{Institute of High Energy Physics, CAS, Beijing, China}
\author{Zhao-Huan Yu}\email{yuzhaoh5@mail.sysu.edu.cn}
\affiliation{School of Physics, Sun Yat-Sen University, Guangzhou 510275, China}

\begin{abstract}

We investigate fermionic dark matter interactions with standard model particles from an additional $\mathrm{U}(1)_\mathrm{X}$ gauge symmetry, assuming kinetic mixing between the $\mathrm{U}(1)_\mathrm{X}$ and $\mathrm{U}(1)_\mathrm{Y}$ gauge fields as well as a nonzero $\mathrm{U}(1)_\mathrm{X}$ charge of the Higgs doublet.
For ensuring gauge-invariant Yukawa interactions and the cancellation of gauge anomalies, the standard model fermions are assigned $Y$-sequential $\mathrm{U}(1)_\mathrm{X}$ charges proportional to the Higgs charge.
Although the Higgs charge should be small due to collider constraints, it is useful to decrease the effective cross section of dark matter scattering off nucleons by two orders of magnitude and easier evade from direct detection bounds.
After some numerical scans performed in the parameter space, we find that the introduction of the Higgs charge can also enhance the dark matter relic density by at least two orders of magnitude. When the observed relic density and the direct detection constraints are tangled, at the case where the resonance effect is important for dark matter freeze-out, the Higgs charge can expand physical windows to some extent by relieving the tension between the relic density and the direct detection.
\end{abstract}

\maketitle
\tableofcontents
\clearpage

\section{Introduction}

The standard model (SM) with $\mathrm{SU}(3)_\mathrm{C} \times \mathrm{SU}(2)_\mathrm{L} \times \mathrm{U}(1)_\mathrm{Y}$ gauge interactions has achieved a dramatic success in explaining experimental data in particle physics.
Nonetheless, the SM must be extended for taking into account dark matter (DM) in the Universe, whose existence is established by astrophysical and cosmological experiments~\cite{Jungman:1995df,Bertone:2004pz,Feng:2010gw,Young:2016ala}.
The standard paradigm for DM production assumes that dark matter is thermally produced in the early Universe, typically requiring some mediators to induce adequate DM interactions with SM particles.

One of the simple attempts is to assume that the DM particle carries a $\mathrm{U}(1)_\mathrm{X}$ charge associated with an additional $\mathrm{U}(1)_\mathrm{X}$ gauge symmetry with the corresponding gauge boson acting as a mediator~\cite{Langacker:2008yv}.
In order to minimize the impact on the interactions of SM particles, one may assume that all SM fields do not carry $\mathrm{U}(1)_\mathrm{X}$ charges~\cite{Feldman:2007wj,Pospelov:2007mp,Mambrini:2010dq,Kang:2010mh,Chun:2010ve,Mambrini:2011dw,Frandsen:2011cg,Gao:2011ka,Chu:2011be,Frandsen:2012rk,Jia:2013lza,Belanger:2013tla,Chen:2014tka,Arcadi:2017kky,Liu:2017lpo,Dutra:2018gmv,Bauer:2018egk,Koren:2019iuv,Jung:2020ukk, ZHHH:2020dm, Cai:2021evx, Liu:2022evb, Carvunis:2022yur, Qiu:2023wbs}.
Thus, the kinetic mixing  between the $\mathrm{U}(1)_\mathrm{X}$ and $\mathrm{U}(1)_\mathrm{Y}$ gauge fields~\cite{Holdom:1985ag,YingZ:2009} induces DM interactions with SM particles.
However, the DM interaction with such a kinetic mixing portal could be too weak to achieve the relic abundance via the freeze-out mechanism~\cite{firstFO,Kolb:1990vq,relicFO}, or too strong to escape from DM direct detection, because a single mixing parameter may be insufficient to satisfy both requirements.
Therefore, it could be useful to introduce more parameters.

%Regarding the fact that direct detection is mainly related by the $\mathrm{U}(1)_\mathrm{X}$ interaction of fermions, one wonders the situation to separate it from the $\mathrm{U}(1)_\mathrm{X}$ interaction of gauge bosons, while the latter may keep DM freeze out as expected.
In this paper, we assume the SM Higgs field also carries a $\mathrm{U}(1)_\mathrm{X}$ charge~\cite{Ekstedt:2016}, which is very small for keeping the new $Z^\prime$ gauge boson weakly coupled to the SM sector.
Because of the kinetic mixing term and the Higgs $\mathrm{U}(1)_\mathrm{X}$ charge, the $\mathrm{U}(1)_\mathrm{X}$ and $\mathrm{U}(1)_\mathrm{Y}$ gauge fields mix with each other, while one electrically neutral gauge boson remains massless, which is the photon. For ensuring the gauge invariance of the SM Yukawa couplings, the SM fermions should also be charged under $\mathrm{U}(1)_\mathrm{X}$. In order to cancel chiral gauge anomalies, we assume that the fermions carry $Y$-sequential $\mathrm{U}(1)_\mathrm{X}$ charges~\cite{Fayet:1989mq,Bogdan:2003,Ekstedt:2016,Fairbairn:2017,Ellis:2018xal}, which are also very small because they should be proportional to the Higgs $\mathrm{U}(1)_\mathrm{X}$ charge.
Such a case is different from those conventionally proposed~\cite{Babu:1997st,Carena:2004,Frandsen:2011cg,adas:2016,Fairbairn:2017,Das:2019pua,Das:2022oyx}, since the latter usually have ${\cal O}(1)$ charges to lift some kind of physical processes. It is also notable that this case is similar to that for the $U$-boson~\cite{Fayet:1990wx,Fayet:2004bw}, in the sense that the $\mathrm{U}(1)_\mathrm{X}$ gauge couplings to SM particles are much weaker than that to dark matter.
%The fermion's $\mathrm{U}(1)_\mathrm{X}$ charges in our work, as a secondary outcome, however are very small since they are proportional to $\mathrm{U}(1)_\mathrm{X}$ charge of Higgs doublet, and the full model are put upon the base of fundamental and complete renormalizable gauge theory.
Now there is one more free parameter, \textit{i.e.,} the Higgs $\mathrm{U}(1)_\mathrm{X}$ charge, to affect the $Z^\prime$ couplings to SM particles.
It is necessary to investigate its compact on DM phenomenology.

In this context, we study a Dirac fermionic DM particle~\cite{Mambrini:2010dq,Chun:2010ve,Liu:2017lpo,Bauer:2018egk}, and observe that the DM couplings to protons and neutrons are typically different~\cite{Kang:2010mh,Frandsen:2011cg,Gao:2011ka,Chun:2010ve,Belanger:2013tla,Chen:2014tka}, leading to isospin-violating DM-nucleon scattering~\cite{Feng:2011vu} in direct detection experiments. It is not obvious whether the correct DM relic abundance can be achieved or not until we carry out some numerical scans.
We find that the presence of the extra parameter can accommodate wider ranges of the $\mathrm{U}(1)_\mathrm{X}$ gauge coupling and the DM particle mass.

This paper is organized as follows.
In Sec.~\ref{sec:U1X_gauge}, we introduce the $\mathrm{U}(1)_\mathrm{X}$ gauge theory where the Higgs doublet carries a $\mathrm{U}(1)_\mathrm{X}$ charge, and discuss the induced interactions of SM fermions. In Sec.~\ref{sec:fermionic_DM}, we study Dirac fermionic DM charged under $\mathrm{U}(1)_\mathrm{X}$, and explore the effective DM-nucleon scattering cross-section for direct detection as well as the DM relic abundance via numerical scans. Finally, we gives the conclusions in Sec.~\ref{sec:concl}.

\section{$\mathrm{U}(1)_\mathrm{X}$ gauge theory}
\label{sec:U1X_gauge}

In this section, we introduce the $\mathrm{U}(1)_\mathrm{X}$ gauge theory with kinetic mixing between the $\mathrm{U}(1)_\mathrm{X}$ and $\mathrm{U}(1)_\mathrm{Y}$ gauge fields. We assign a small $\mathrm{U}(1)_\mathrm{X}$ charge to the SM Higgs doublet and the SM Yukawa interactions are gauge-invariant only if the SM fermions have appropriate $\mathrm{U}(1)_\mathrm{X}$ charges, which are chosen to be $Y$-sequential, \textit{i.e.}, obey the same relations as their $\mathrm{U}(1)_\mathrm{Y}$ charges, so that the theory remains free from chiral anomalies.

\subsection{$\mathrm{U}(1)_\mathrm{X}$ gauge theory with $\mathrm{U}(1)_\mathrm{X}$-charged SM Higgs doublet}

We denote the $\mathrm{U}(1)_\mathrm{Y}$ and $\mathrm{U}(1)_\mathrm{X}$ gauge fields as $\hat{B}_\mu$ and $\hat{Z}'_\mu$, respectively.
Their gauge-invariant kinetic terms in the Lagrangian read
\begin{equation}\label{kin_mix}
\mathcal{L}_{\mathrm{K}} =
-\frac{1}{4} \hat{B}^{\mu\nu}\hat{B}_{\mu\nu}
-\frac{1}{4} \hat{Z}'^{\mu\nu}\hat{Z}'_{\mu\nu}
-\frac{\sin \epsilon}{2} \hat{B}^{\mu\nu}\hat{Z}'_{\mu\nu} ,
\end{equation}
where the field strengths are $\hat{B}_{\mu\nu} \equiv \partial_\mu \hat{B}_\nu - \partial_\nu \hat{B}_\mu$ and $\hat{Z}'_{\mu\nu} \equiv \partial_\mu \hat{Z}'_\nu - \partial_\nu \hat{Z}'_\mu$.
The $\sin\epsilon$ term is a kinetic mixing term, which makes the kinetic Lagrangian \eqref{kin_mix} in a noncanonical form.

We assume that the $\mathrm{U}(1)_\mathrm{X}$ gauge symmetry is spontaneously broken \cite{Higgs:1964ia,Higgs:1964pj,Englert:1964et} by a Higgs field $\hat{S}$ with $\mathrm{U}(1)_\mathrm{X}$ charge $x_S = 1$\footnote{The Stueckelberg mechanism~\cite{Stueckelberg:1938hvi,Chodos:1971yj} is another way to generate the $\mathrm{U}(1)_\mathrm{X}$ gauge boson mass.}.
Now the Higgs sector involves $\hat{S}$ and the SM Higgs doublet $\hat{H}$. The corresponding Lagrangian respecting the $\mathrm{SU}(2)_\mathrm{L} \times \mathrm{U}(1)_\mathrm{Y} \times \mathrm{U}(1)_\mathrm{X}$ gauge symmetry is~\cite{Liu:2017lpo}
\begin{eqnarray}\label{modelLagarange}
\mathcal{L}_{\mathrm{H}} &=& (D^\mu \hat{H})^\dag(D_\mu \hat{H})+(D^\mu \hat{S})^\dag(D_\mu \hat{S}) +\mu^2|\hat{H}|^2 +\mu_S^2
|\hat{S}|^2
\nonumber\\
&& -\frac{1}{2}\lambda_H|\hat{H}|^4 -\frac{1}{2}\lambda_S|\hat{S}|^4 -\lambda_{HS}|\hat{H}|^2|\hat{S}|^2.
\end{eqnarray}
The covariant derivatives are given by 
\begin{eqnarray}
D_\mu\hat{H}&=&(\partial_\mu-i Y_H \hat{g}'\hat{B}_\mu -i\zeta g_{X}\hat{Z}'_\mu  -i\hat{g} W^a_\mu T^a)\hat{H}, \\  
D_\mu\hat{S}&=&(\partial_\mu-i g_{X} \hat{Z}'_\mu - i Y_S \hat{g}'\hat{B} )\hat{S},
\end{eqnarray}
where $W^a_\mu$ ($a=1,2,3$) denote the $\mathrm{SU}(2)_\mathrm{L}$ gauge fields and $T^a = \sigma^a/2$ are the $\mathrm{SU}(2)_\mathrm{L}$ generators.
$\hat{g}$, $\hat{g}'$, and $g_X$ are the $\mathrm{SU}(2)_\mathrm{L}$, $\mathrm{U}(1)_\mathrm{Y}$, $\mathrm{U}(1)_\mathrm{X}$ gauge couplings, respectively. The hypercharge $Y_H = 1/2$ for $\hat{H}$ is the same as in the SM.

The presence of $\zeta$ and $Y_S$ terms is notable here. They generally reflect a $\mathrm{U}(1)_\mathrm{X}$ charge of the SM Higgs doublet $\hat{H}$ and the $\mathrm{U}(1)_\mathrm{Y}$ charge of the exotic Higgs field $\hat{S}$.
Some studies thought this $\zeta$ charge can be absorbed into $g_X$ by a scaling, but in this work it is found to be an independent parameter. It will predict a different phenomenology as shown in the following. Before going to any detailed analysis, it is also necessary to point out that, 
in comparison to the Higgs charges introduced in Refs.~\cite{Carena:2004,Frandsen:2011cg,adas:2016,Fairbairn:2017} which were usually $\sim {\cal O}(1)$, the magnitude of $\zeta$ in this work is expected to be very small, such that ${\hat Z}^\prime$ would have a weak connection to SM particles. Nonetheless, compared to the size of the kinetic mixing parameter $\sin\epsilon$, the value of $\zeta g_X$ is not necessarily smaller. In fact, it is invited to balance the effect from the former.

Both $\hat{H}$ and $\hat{S}$ acquire nonzero vacuum expectation values (VEVs), $v$ and $v_S$, driving the spontaneously symmetry breaking of gauge symmetries.
The Higgs fields in the unitary gauge can be expressed as
\begin{eqnarray}
\hat{H}&=&
\frac{1}{\sqrt{2}}
\begin{pmatrix}
0\\
v+H
\end{pmatrix},
\\
\hat{S}&=&
\frac{1}{\sqrt{2}} (v_S+S).
\end{eqnarray}
Vacuum stability requires the following conditions:
\begin{eqnarray}
\lambda _H > 0,\quad
\lambda_S > 0,\quad
\lambda_{HS} > -\sqrt{\lambda_H\lambda_S}\,.
\end{eqnarray}
There is a transformation from the gauge basis $(H,S)$ to the mass basis $(h,s)$,
\begin{equation}
\label{eq:HS_to_hs}
\begin{pmatrix}
H\\
S
\end{pmatrix} =
\begin{pmatrix}c_\eta&-s_\eta\\
s_\eta&c_\eta
\end{pmatrix}  
\begin{pmatrix}
h\\
s
\end{pmatrix},
~~~  
\tan{2\eta}=\frac{2\lambda_{HS}v v_S}{\lambda_Hv^2-\lambda_Sv_S^2},
\end{equation}
with the mixing angle $\eta \in [-\pi/4,\pi/4]$. The physical eigenstate $h$ is the $125~\si{GeV}$ SM-like Higgs boson, whose properties would be identical to the SM Higgs boson if $\lambda_{HS}$ and $\zeta$ vanish. The exotic Higgs boson $s$ can be assumed to be very heavy and give no effect on TeV phenomena. 

The mass-squared matrix for the gauge fields $(\hat{B}_\mu, W^3_\mu, \hat{Z}'_\mu)$ generated by the Higgs VEVs reads
\begin{equation}
\mathcal{M}_{VV^{\prime}}^2 =
\begin{pmatrix}
\hat{g}'^2 ( v^2/4 + Y_S^2 v^2_S) ~~ & -\hat{g}\hat{g}' v^2/4   ~~&  \mathcal{M}_{13}^2    \\
- \hat{g}\hat{g}'v^2 /4         &\hat{g}^2 v^2/4  & \mathcal{M}_{23}^2   \\
\mathcal{M}_{13}^2  & \mathcal{M}_{23}^2   & g_X^2 ( \zeta^2 v^2 + v_S^2 )  
\end{pmatrix},
\end{equation} 
with
\begin{equation} 
 \mathcal{M}_{13}^2 = \hat{g}' g_X ( Y_S v_S^2  + \zeta v^2 /2 ),\quad
 \mathcal{M}_{23}^2 =  - \zeta \hat{g} g_X v^2 /2.
\end{equation}
This can be regarded as a generalization of the simplest Higgs structure realized in Ref.~\cite{Babu:1997st}.
Note that $\mathcal{M}_{23}^2$ is present only for $\zeta \neq 0$.
The transformation from the gauge basis $(\hat{B}_\mu, W^3_\mu, \hat{Z}'_\mu)$  to the mass basis $(A_\mu, Z_\mu, Z'_\mu)$ can be expressed as~\cite{Frandsen:2011cg}
\begin{equation}\label{rotation}
\begin{pmatrix}
\hat{B}_\mu\\
W^3_\mu\\ \hat{Z}'_\mu
\end{pmatrix}
=
 V(\epsilon ) R_3(\hat{\theta}_\mathrm{W}) R_1(\xi)
\begin{pmatrix}
A_\mu\\
Z_\mu\\
Z'_\mu
\end{pmatrix},
\end{equation}
with
\begin{equation}
V(\epsilon ) = 
\begin{pmatrix}
1 & & - t_{\epsilon}\\
  & 1 & \\
0 & & \frac{1}{c_\epsilon} 
\end{pmatrix} , \quad
R_3(\hat{\theta}_\mathrm{W})= 
\begin{pmatrix}
\hat{c}_\mathrm{W}&-\hat{s}_\mathrm{W}& \\
\hat{s}_\mathrm{W}&\hat{c}_\mathrm{W} \\
 & &1\end{pmatrix}, \quad
R_1(\xi)=
\begin{pmatrix}
1& & \\
 &c_\xi&-s_\xi\\
 &s_\xi&c_\xi
\end{pmatrix},
\end{equation}
to make the kinetic terms canonical and the mass-squared matrix diagonalized\footnote{Through this text, $s$, $c$, and $t$ denote the sine, cosine, and tangent functions, with the subscript denoting the argument. In particular, we define $\hat{s}_\mathrm{W} \equiv \sin\hat{\theta}_\mathrm{W}$ and $\hat{c}_\mathrm{W} \equiv 
\cos\hat{\theta}_\mathrm{W}$.}. 
$V(\epsilon)$ is a 3-dimensional extension to a $\mathrm{GL}(2,\mathbb{R})$ transformation among $(\hat{B}_\mu, \hat{Z}'_\mu)$~\cite{Babu:1997st}, which makes the kinetic Lagrangian~\eqref{kin_mix} canonical. The kinetic mixing parameter $\epsilon$ should satisfy $\epsilon \in (-1,1)  $ to ensure correct signs for the canonical kinetic terms.
Note that the $A_\mu$ and $Z_\mu$ field correspond to the photon and the $Z$ boson, and the $Z'_\mu$ field leads to a new neutral massive vector boson $Z'$. After the mass diagonalization, the photon mass squared reads
\begin{eqnarray}
M^2_{\gamma\gamma} &=& \frac{1}{4}[ \hat{g}'^2 ( v^2 + 4 Y_S^2 v^2_S ) \hat{c}_\mathrm{W}^2 - 2  \hat{g}' \hat{g} v^2 \hat{c}_\mathrm{W} \hat{s}_\mathrm{W} + \hat{g}^2 v^2 \hat{s}_\mathrm{W}^2 ]  \\  \nonumber
& = & Y_S^2 v^2_S \hat{c}_\mathrm{W}^2 + \frac{1}{4}( \hat{g}' \hat{c}_\mathrm{W}  - \hat{g} \hat{s}_\mathrm{W})^2 v^2.
\end{eqnarray}
In order to keep the photon massless, it is natural for $Y_S$ and the weak mixing angle $\hat\theta_W$ to satisfy
\begin{equation}
Y_S = 0,  \quad
\hat{s}_\mathrm{W } = \frac{\hat{g}'}{\sqrt{\hat{g}^2 + \hat{g}'^2}},\quad
\hat{c}_\mathrm{W } = \frac{\hat{g}}{\sqrt{\hat{g}^2 + \hat{g}'^2}}.
\end{equation}
Thus, the exotic Higgs field $\hat{S}$ cannot be charged under $\mathrm{U}(1)_\mathrm{Y}$.
Furthermore, the vanishing of the $Z$-$Z'$ mass term $M^2_{Z Z^{\prime}}$ determine the rotation angle $\xi$ to be\footnote{It is easy to find this
 $t_{2\xi} = \frac{ 2 \sqrt{\hat{g}^2 + \hat{g}^{\prime 2} }  (  s_{\epsilon}  \hat{g}^{\prime} - 2 \zeta  g_X ) v^2 }
{ [ (\hat{g}^2 + \hat{g}^{\prime 2} ) v^2 c^2_\epsilon + 4 g_{X}^{2} ( v_{S}^{2} + \zeta v^2 ) + 4 \zeta s_\epsilon g_X \hat{g}^{\prime} v^2 - s^2_{\epsilon} \hat{g}^{\prime 2} v^2 ]/c_\epsilon } $ 
 is identical to Eq.~(19) in Ref.~\cite{ZHHH:2020dm} for $\zeta = 0$.
This equation has an uninterested solution of $t_{2\xi}= 0$ at $\zeta=g^\prime s_\epsilon /( 2 g_X ) $, which will be ignored hereafter since it will lead to the vanishing of the $Z^\prime$ couplings to SM fermions.}
\begin{equation}\label{defXi}
t_{2 \xi} \equiv \tan 2\xi = 
\frac{ 2 {\cal Z } \hat{s}_\mathrm{W } }{ 1 - r  - ( 1 + r ) C_Z }
\end{equation}
with
\begin{equation}
{\cal Z} \equiv   t_{\epsilon} - \frac{2 \zeta g_X}{\hat{g}^\prime c_\epsilon}, ~~  
r \equiv \frac{m^2_{Z^\prime}}{m^2_Z}.
\end{equation}
Therein $m_{Z^{\prime}}$ and $m_{Z}$ are the physical masses of the vector bosons $Z^\prime$ and $Z$, respectively, while $C_Z$ is a small correction originated from nonvanishing $\epsilon$ and $\zeta$.
The details are given in the appendix. It is notable that because of the existence of $\zeta$, such a mixing represented by the angle $\xi$ does not vanish in the limit $\epsilon \rightarrow 0 $.

\subsection{SM Fermions under  $\mathrm{U}(1)_\mathrm{X}$}

Because the Higgs doublet $\hat{H}$ carries a $\mathrm{U}(1)_\mathrm{X}$ charge $\zeta$, the SM fermions should also have appropriate $\mathrm{U}(1)_\mathrm{X}$ charges to keep the SM Yukawa couplings respecting the $\mathrm{U}(1)_\mathrm{X}$ gauge symmetry.
Thus, the covariant derivatives of the SM quark fields in the gauge basis can be expressed as
\begin{eqnarray}
D_\mu \left( \begin{array}{c} u'_{i\mathrm{L}} \\  d'_{i\mathrm{L}}  \end{array} \right)  &=& [ \partial_\mu- i ( Y_q \hat{g}'\hat{B}_\mu + \hat{g} W^a_\mu \tau^a +  x^\mathrm{L}_q g_{X}\hat{Z}'_\mu ) ] \left( \begin{array}{c} u'_{i\mathrm{L}} \\  d'_{i\mathrm{L}}  \end{array} \right),
 \\  
D_\mu u'_{i\mathrm{R}} &=& [ \partial_\mu- i ( Y_u \hat{g}'\hat{B}_\mu  + x^\mathrm{R}_u g_{X}\hat{Z}'_\mu ) ] u'_{i\mathrm{R}},
\\
D_\mu d'_{i\mathrm{R}} &=&[ \partial_\mu- i ( Y_d \hat{g}'\hat{B}_\mu  + x^\mathrm{R}_d g_{X}\hat{Z}'_\mu ) ] d'_{i\mathrm{R}},
\end{eqnarray}
where $i=1,2,3$ is the generation index.
Therein $x^\mathrm{L}_q$, $x^\mathrm{R}_u$, and $x^\mathrm{R}_d$ are the  $\mathrm{U}(1)_\mathrm{X}$ charges of the left-handed quark doublet, the right-handed up-type quark singlet, and the left-handed down-type quark singlet, respectively. $Y_{q,u,d}$ is the $\mathrm{U}(1)_\mathrm{Y}$ hypercharges as in the SM.

On a necessary condition
\begin{equation}
\zeta = x^\mathrm{L}_q - x^\mathrm{R}_d = x^\mathrm{R}_u - x^\mathrm{L}_q 
\end{equation}
the SM Yukawa interactions of quarks and the Higgs doublet respect the $\mathrm{U}(1)_\mathrm{X}$ gauge symmetry.
For SM leptons, a similar argument leads to $\zeta = x^\mathrm{L}_l - x^\mathrm{R}_l$.
On the other side, in order to cancel the chiral anomalies, all these $\mathrm{U}(1)_\mathrm{X}$ charges are further bounded. 
In this work, we make a simple choice to assume the $\mathrm{U}(1)_\mathrm{X}$ charges of SM fermions proportional to their $\mathrm{U}(1)_\mathrm{Y}$ charges.
This is the so-called $Y$-sequential charges~\cite{Bogdan:2003}, as listed in Tab.~\ref{uxferm}.
\begin{table}
\begin{center}
\setlength\tabcolsep{.6em}
\renewcommand{\arraystretch}{1.5}
\caption{$Y$-sequential $\mathrm{U}(1)_\mathrm{X}$ charges for SM fermions in the gauge basis.}
\begin{tabular}{  c | c  c  c c c }
\hline
Fermions & $u'_{i\mathrm{L}},d'_{i\mathrm{L}}$  &  $u'_{i\mathrm{R}}$  &  $d'_{i\mathrm{R}}$  & $l'_{i\mathrm{L}}, \nu'_{i\mathrm{L}}$ & $l'_{i\mathrm{R}}$  \\
\hline
$\mathrm{U}(1)_\mathrm{X}$ charges $x_f^{\mathrm{L},\mathrm{R}}$    & $\zeta/3$~ &  ~$4\zeta/3$~  & ~$-2\zeta/3$~  &  ~$-\zeta$~  & ~$-2\zeta$     \\
\hline   
\end{tabular}
\label{uxferm}
\end{center}
\end{table}

The charge current interactions of SM fermions at tree level are not affected by the kinetic or mass mixing, keeping the SM form of
\begin{equation}
{\mathcal{L}_{{\mathrm{CC}}}} = \frac{1}{{\sqrt 2 }}(W_\mu ^ + J_W^{ + ,\mu } + \text{H.c.}),
\end{equation}
where the charge current is $J_W^{ + ,\mu } = {\hat g} ( {{\bar u}_{i{\mathrm{L}}}}{\gamma ^\mu }{V_{ij}}{d_{j{\mathrm{L}}}} + {{\bar \nu }_{i{\mathrm{L}}}}{\gamma ^\mu }{\ell _{i{\mathrm{L}}}})$ and $V_{ij}$ is the Cabibbo-Kobayashi-Maskawa matrix.

The neutral current interactions are given by
\begin{equation}
{\mathcal{L}_{{\mathrm{NC}}}} = j_{{\mathrm{EM}}}^\mu {A_\mu } + j_Z^\mu {Z_\mu } + j_{Z'}^\mu {Z'_\mu }.
\end{equation}
Here $j_\mathrm{EM}^\mu  = \sum_f {{Q_f}e\bar f{\gamma ^\mu }f}$ the electromagnetic current with $e \equiv \hat{g} \hat{g}' /\sqrt{\hat{g}^2 + \hat{g}^{'2}}$. $Q_f$ is the electric charge of a fermion $f$ in the mass basis.
The $Z$ neutral current is
\begin{eqnarray}\label{j_Z_mu}
j_Z^\mu  &=& \frac{e {\tilde c}^+_\xi}{{2{{\hat s}_{\mathrm{W}}}{{\hat c}_{\mathrm{W}}}}}\sum\limits_f {\bar f{\gamma ^\mu }(  T_f^3 - 2{Q_f} s_*^2 - T_f^3{\gamma _5}  f}    \nonumber \\
&& + \frac{g_X s_\xi }{2 c_\epsilon } \sum\limits_f { ( x^\mathrm{L}_f + x^\mathrm{R}_f ) \bar f {\gamma ^\mu} f  }
+ \frac{g_X s_\xi }{2 c_\epsilon } \sum\limits_f { ( x^\mathrm{R}_f - x^\mathrm{L}_f ) {\bar f {\gamma ^\mu} {\gamma_5 f }} }+ \frac{s_\xi}{c_\epsilon}  j^\mu_\mathrm{DM},
\end{eqnarray}
with $T^3_f$ corresponding to the third component of the weak isospin of $f$
and 
\begin{equation}
{\tilde c}^\pm_\xi \equiv c_\xi \pm {\hat s}_\mathrm{W} t_\epsilon s_\xi,\quad
s_*^2 =  \hat s_{\mathrm{W}}^2 + \frac{\hat c_{\mathrm{W}}^2{\hat s}_{\mathrm{W}}{t_\epsilon }{t_\xi }}{{1 + {{\hat s}_{\mathrm{W}}}{t_\epsilon }{t_\xi }}}.
\end{equation}
The $Z'$ neutral current is
\begin{equation}\label{j_Zprime_mu}
j_{Z^\prime}^\mu  =  
\sum\limits_f {\bar f{\gamma ^\mu }( v_f + a_f {\gamma _5} )f} 
+ \frac{ c_\xi }{c_\epsilon }  j^\mu_\mathrm{DM},
\end{equation}
with
\begin{eqnarray}
v_f &=& -\frac{e {\tilde s}^-_\xi (T_f^3 - 2 Q_f \hat s^2_{\mathrm{W}})}{ 2 {\hat s}_{\mathrm{W}} {\hat c}_{\mathrm{W}}}
  -  Q_f e {\hat c}_{\mathrm{W}} t_\epsilon c_\xi + \frac{g_X c_\xi ( x^L_f + x^R_f ) }{2 c_\epsilon },
\\
a_f &=&  \frac{e {\tilde s}^-_\xi T_f^3}{ 2 {\hat s}_{\mathrm{W}} {\hat c}_{\mathrm{W}}}  + \frac{g_X c_\xi ( x^R_f - x^L_f ) }{2 c_\epsilon },
\qquad  {\tilde s}^\pm_\xi \equiv  s_\xi \pm {\hat s}_\mathrm{W} t_\epsilon c_\xi .
\end{eqnarray}
It is remarkable that, at the limit  $ \epsilon \rightarrow 0 $, the corrections to the interactions between the SM fermions and the $Z$ boson will be proportional to $\zeta$, just like their couplings to $Z^{\prime}$.
%\footnote{In some literatures the $\mathrm{U}(1)_\mathrm{X}$ charges are supposed to be rational numbers, this is feasible in this work even though $\zeta$ will be taken as very small at the similar order of $\epsilon$.}
Recall that $t_\xi$ (and hence ${\tilde s}^\pm_\xi$ and ${\tilde c}^\pm_\xi$ ) implicitly depends on $\zeta$, so Eqs.~(\ref{j_Z_mu}) and (\ref{j_Zprime_mu}) explicitly demonstrate that $\zeta$ cannot be absorbed into a redefinition of $g_X$.
 
\subsection{ Parameterization and constraints}

The discussions in the above subsections indicate that not all the presented parameters are independent. It is necessary to define a convenient scheme for later calculation. Firstly, the photon couplings to SM fermions remain the same forms as in the SM at tree level,
with the electric charge unit $e=\sqrt{4 \pi \alpha}$ can be determined by the $\overline{\mathrm{MS}}$ fine-structure constant $\alpha(m_Z) = 1/127.955$ at the $Z$ pole~\cite{weinbergAngle}.
The mass of the $W$ boson receive contribution only from the Higgs doublet VEV $v$ in the form $m_W = {\hat g} v/2$, leading to an expression of $v$ from the Fermi constant $G_\mathrm{F} = \hat{g}^2/(4\sqrt{2}m_W^2) =(\sqrt{2}v^2)^{-1}$.

The electroweak gauge couplings $\hat{g}$ and $\hat{g}'$ are related to $e$ through $\hat{g}=e/\hat{s}_\mathrm{W}$ and $\hat{g}'=e/\hat{c}_\mathrm{W}$, but the Weinberg angle $\hat{\theta}_\mathrm{W}$ is corrected by new physics.
In the $\mathrm{U}(1)_\mathrm{X}$ gauge theory, it is straightforward to get a relation at tree level,
\begin{equation}
\hat{s}_\mathrm{W}^2 \hat{c}_\mathrm{W}^2 = \frac{{\pi \alpha }}{{\sqrt 2 {G_\mathrm{F}} \hat{m}_Z^2}}.
\end{equation}
Comparing to its SM counterpart $s_\mathrm{W}^2c_\mathrm{W}^2 = \pi \alpha /( \sqrt 2 {G_\mathrm{F}}m_Z^2 )$ and
utilizing Eq.~\eqref{zsmass} in the appendix, we have
\begin{equation}\label{weak_mixing_angle}
s_\mathrm{W}^2c_\mathrm{W}^2
= \frac{\hat{s}_\mathrm{W}^2\hat{c}_\mathrm{W}^2}{1+ C_Z },
\end{equation}
where $C_Z$ is defined in Eq.~\eqref{cZZp}.
Therefore, the hatted weak mixing angle $\hat\theta_\mathrm{W}$ can be expressed as a correction added to its SM counterpart, while the latter   
are determined by the best-measured parameters $\alpha$, $G_\mathrm{F}$, and $m_Z$~\cite{weinbergAngle,Burgess:1993vc}.

The rotation angle $\xi$ can be represented as a function of fundamental parameters like $g_X$, $m_{Z'}$, $\epsilon$, and $\zeta$. With the procedure described in the appendix, one can find an approximate solution as
\begin{equation}
t_{2\xi} = \frac{2 {\cal Z} s_\mathrm{W} }{ 1- r } 
-  \frac{2 (1+r) {\cal Z}^3 s^3_\mathrm{W} }{{(1-r)}^3} 
+  \frac{ {\cal Z}^2 s^3_\mathrm{W} c^2_\mathrm{W} }{  ( c^2_\mathrm{W} - s^2_\mathrm{W} ) {( 1- r )}^2 }\left( {\cal Z} + \zeta \frac{ g_X}{ \sqrt{\pi \alpha} }\frac{s^2_\mathrm{W}}{c_\mathrm{W} c_\epsilon} \right).
\label{prxmtXi}
\end{equation}
From this equation, one can inversely solve $\zeta$ as a function of $t_\xi$.
Thus, $t_\xi$ can be regarded as a free parameter, while $\zeta$ becomes a induced parameter. Fortunately, the procedure can be traded in an exact way as detailed in the appendix. It is obvious that $t_\xi$ is of more convenience as a free parameter for phenomenological discussions. Hereafter, we adopt a free parameter set as
\begin{equation}
\{g_X,~ m_{Z'},~  t_{\epsilon},~  t_\xi, ~ m_s, ~ s_{\eta}\}.
\end{equation}
From these free parameters, we can derive all other parameters based on the above expressions\footnote{The relations between the free and induced parameters in the Higgs sector are further described in the appendix.}.

These free parameters are constrained by the measurements of the $Z{\bar f} f$ vector and axial-vector couplings, where the LEP-II precise measurements is most important. The quantities like $\Gamma_Z$, $A^{(0,e)}_{FB}$, $A^{(0,c)}_{FB}$, and $A^{(0,b)}_{FB}$\footnote{Regarding that the universality among three generations is still kept.} will be recalculated in our model and confirmed within the experimental limits from Tab.~10.5 in Ref.~\cite{weinbergAngle}. The measurements at the $Z$ pole further require that the correction to the Weinberg angle $s^2_\mathrm{W}$ need to be sufficiently small, rendering the couplings of gauge bosons close to their SM values. Moreover, the searches for the $Z^\prime$ boson at the LHC~\cite{ATLAS:2019erb,CMS:2021ctt} have put constraints on the $Z^\prime {\bar f} f$ couplings. The mixing angle $\eta$ between the two Higgs bosons will be set sufficiently small ($\le 0.1$), so no deviation is expected in the Higgs phenomena.   

\section{Dirac fermionic dark matter}
\label{sec:fermionic_DM}

We are interested in the connection between the $Z^\prime$ boson and dark matter phenomenology. In this section, we discuss the case that the DM particle is a Dirac fermion $\chi$ with a $\mathrm{U}(1)_\mathrm{X}$ charge $q_\chi$~\cite{Mambrini:2010dq,Chun:2010ve,Liu:2017lpo,Bauer:2018egk}.
The Lagrangian for $\chi$ reads
\begin{eqnarray}
 \mathcal{L}_{\chi}=i \bar{\chi} \gamma^{\mu} D_{\mu} \chi-m_{\chi} \bar{\chi} \chi,
 \end{eqnarray}
where $D_\mu \chi  =(\partial_\mu - i q_\chi g_X \hat{Z}'_\mu)\chi$ and $m_\chi$ is the $\chi$ mass.
Thus, the DM neutral current appearing in Eqs.~\eqref{j_Z_mu} and \eqref{j_Zprime_mu} is
\begin{equation}
j^\mu_\mathrm{DM} = q_\chi g_X \bar\chi \gamma^\mu \chi.
\end{equation}
Thus, the $Z$ and $Z'$ bosons mediate the interaction between DM and SM fermions.
The number densities of $\chi$ and its antiparticle $\bar\chi$ yielded by the freeze-out mechanism should be equal. Both $\chi$ and $\bar\chi$ fermions make up dark matter in the Universe.
Below, we study the phenomenology of DM direct detection, as well as relic abundance and indirect detection.
$q_\chi = 1$ will be adopted in the following calculation.

\subsection{Direct detection}

Only the  vector current interactions between $\chi$ and quarks contribute to DM scattering off nuclei in the zero momentum transfer limit, at which the DM direct detection experiments essentially operate.
In the context of effective field theory~\cite{Zheng:2010js}, the interactions between the DM fermion $\chi$ and SM quarks $q$ can be described by
\begin{equation}\label{DM-q_scat}
\mathcal{L}_{\chi q}=\sum_{q} G^{\mathrm{V}}_{\chi q} \bar{\chi} \gamma^{\mu} \chi \bar{q} \gamma_{\mu} q,
\end{equation}
with
\begin{equation}\label{G_V_chiq}
G^{\mathrm{V}}_{\chi q} = -\frac{q_{\chi} g_{X}}{c_{\epsilon}}\left(\frac{s_{\xi} g_{Z}^{q}}{m_{Z}^{2}}+\frac{c_{\xi} g_{Z^{\prime}}^{q}}{m_{Z^{\prime}}^{2}}\right).
\end{equation}
From Eqs.~\eqref{j_Z_mu} and \eqref{j_Zprime_mu}, the vector current couplings of quarks to the $Z$ and $Z'$ bosons are given by
\begin{equation}
g_{Z}^{q} = \frac{{e{c_\xi }}(1 + {{\hat s}_{\mathrm{W}}}{s_\epsilon }{t_\xi })}{{2{{\hat s}_{\mathrm{W}}}{{\hat c}_{\mathrm{W}}}}}(T_q^3 - 2{Q_q}s_*^2),
~~~  g_{Z^{\prime}}^{q} = v_q.
\label{g_Zprime_q}
\end{equation}

The effective Lagrangian for DM-nucleon interactions induced by the DM-quark interactions is
\begin{eqnarray}
\mathcal{L}_{\chi N}=\sum_{N =p,n} G^\mathrm{V}_{\chi N} \bar{\chi} \gamma^{\mu} \chi \bar{N} \gamma_{\mu} N,
\end{eqnarray}
where $G^\mathrm{V}_{\chi p}=2 G^\mathrm{V}_{\chi u}+G^\mathrm{V}_{\chi d}$
and $G^\mathrm{V}_{\chi n} = G^\mathrm{V}_{\chi u} + 2 G^\mathrm{V}_{\chi d}$ counts the contributions of valence quarks to the vector current interactions of nucleons.
Following the strategy in Refs.~\cite{Feng:2011vu,Feng:2013fyw,ZHHH:2020dm}, the effective spin-independent (SI) DM-nucleon cross section for isotope nuclei with atomic number $Z$ can be recast as
\begin{equation}
\sigma^\mathrm{SI}_{\chi N} = {\sigma _{\chi p}}\,\frac{{\sum_i {{\eta _i}\mu _{\chi {A_i}}^2{{[Z + ({A_i} - Z){G^\mathrm{V}_{\chi n}}/{G^\mathrm{V}_{\chi p}}]}^2}} }}{{\sum_i {{\eta _i}\mu _{\chi {A_i}}^2A_i^2} }},
\label{xschiN}
\end{equation}
where $\sigma _{\chi p}$ is the DM-proton scattering cross section.
$\mu_{\chi A_i} \equiv m_\chi m_{A_i}/(m_\chi + m_{A_i})$ is the reduced mass of $\chi$ and a isotope nucleus with mass number $A_i$ and fractional number abundance $\eta_i$.
We use this expression to compare the model prediction to the experimental results expressed by the normalized-to-nucleon cross section.

\begin{figure}[!t]
	\centering
	{\includegraphics[height=0.35\textheight,width=0.6\textwidth]{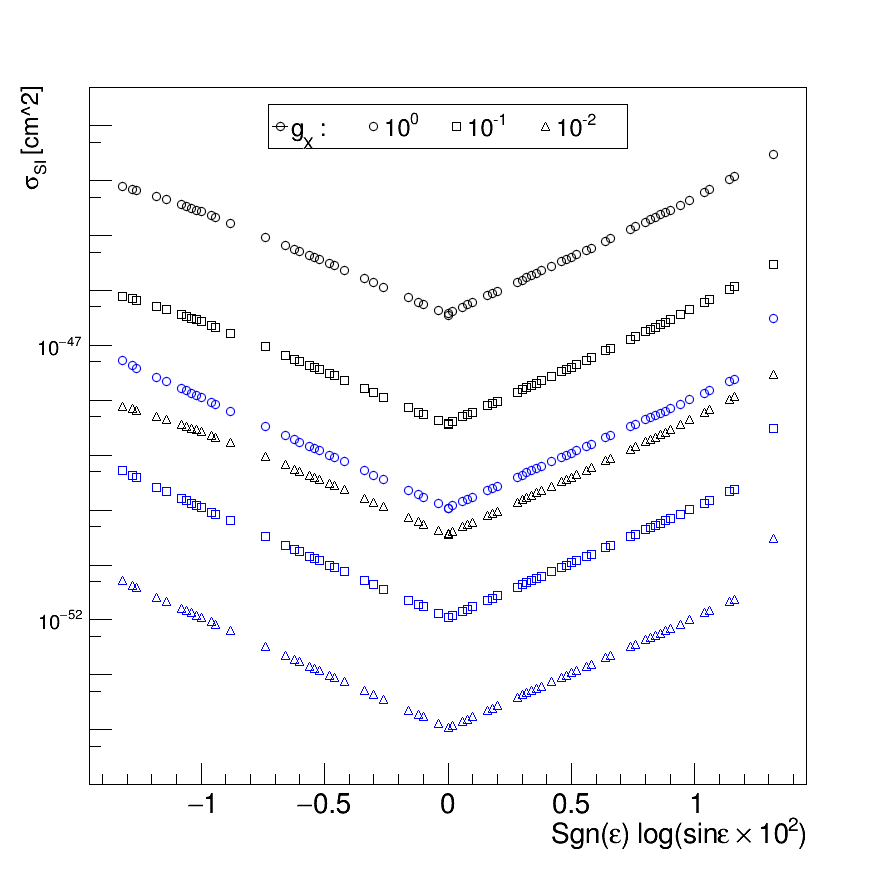}}
	\caption{$\sigma^\mathrm{SI}_{\chi N}$ dependence on $\sin\epsilon$ for $m_\chi = 120~\si{GeV}$, $m_{Z^\prime}= 500~\si{GeV}$, and $g_X = 0.01$ (triangles), $0.1$ (squares), and $1$ (circles). The zero on the horizontal coordinate means $\sin\varepsilon = {10}^{-2}$, while $\pm 1$ means $\sin\varepsilon = \pm {10}^{-1}$. The black points correspond to $\zeta=0$. The blue points are derived by adjusting $\zeta$ to achieve a cancellation in $\sigma^\mathrm{SI}_{\chi N}$.}
\label{isospin}
\end{figure}

Such a setup typically leads to isospin violation in DM-nucleon scatterings. The case of $\zeta = 0$ gives ${G^\mathrm{V}_{\chi n}} = 0 \neq G^\mathrm{V}_{\chi p}$~\cite{ZHHH:2020dm}. In the case of a nonzero $\zeta$, however, we find that ${G^\mathrm{V}_{\chi n}} = 0$ is no longer held. What is more interesting is that the presence of $\zeta$ is able to bring us a relative minus sign between the neutron coupling ${G^\mathrm{V}_{\chi n}}$ and the proton coupling ${G^\mathrm{V}_{\chi p}}$. Eventually, a nonzero $\zeta$ may lead to destructive interference in the total cross section, which may help the model survive from the stringent direct detection constraints.

In Fig.~\ref{isospin}, we show the $\sigma^\mathrm{SI}_{\chi N}$ dependence on $\sin\epsilon$ for $g_X = 0.01, 0.1, 1$ assuming liquid xenon as detection material with $m_\chi = 120~\si{GeV}$ and $m_{Z^\prime}= 500~\si{GeV}$ fixed.
The black points correspond to $\zeta=0$, while the blue points are given by adjusting $\zeta$ for each $\sin\epsilon$ to achieve a cancellation in $\sigma^\mathrm{SI}_{\chi N}$.
The calculation is double-checked by both the formula and the \texttt{MadDM} code~\cite{Backovic:2013dpa, Ambrogi:2018jqj}.
It is easily observed that $\sigma^\mathrm{SI}_{\chi N}$ can be decreased by two orders of magnitude for appropriate $\zeta$. Thus, this model could easily survive in the recent direct detection experiments~\cite{PandaX-4T:2021bab, LUX-ZEPLIN:2022xrq, XENON:2023cxc}.

\subsection{Relic abundance and numerical scan}

\begin{figure}[!t]
	\centering
	\includegraphics[height=0.35\textheight,width=0.6\textwidth]{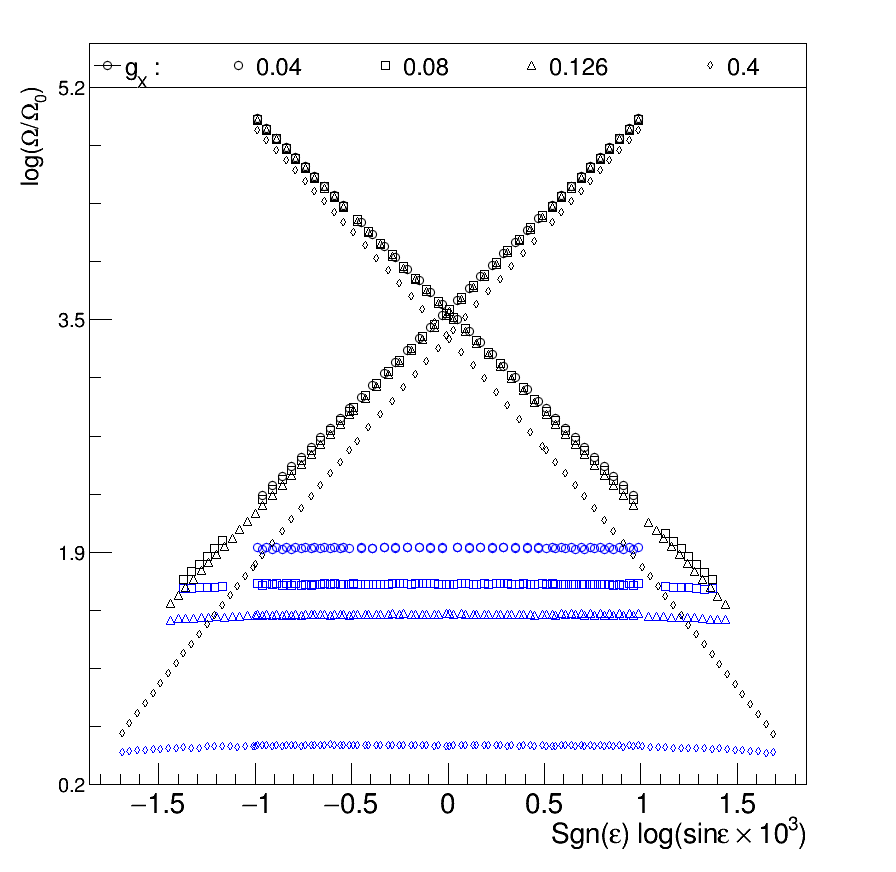}
	\caption{The DM relic density expressed as $\log(\Omega/\Omega_0)$ for nonzero $\zeta$ (blue points) and $\zeta = 0$ (black points) with $m_\chi = 210~\si{GeV}$,  $m_{Z^\prime}= 500~\si{GeV}$, and $g_X = 0.04$ (circles), $0.08$ (squares), $0.126$ (triangles), and $0.4$ (diamonds). The zero on the vertical coordinate means $\Omega=\Omega_0$. The zero on the horizontal coordinate means $\sin\varepsilon  = {10}^{-3}$, while $\pm 1$ means $ \sin\varepsilon = \pm {10}^{-1}$.}
\label{relicenhanced}
\end{figure}

%In the early Universe, $\chi$ and $\bar\chi$ particles would be produced either from FIMP decays, or be annihilated sufficiently quick into SM particles so that the total DM relic abundance should be consistent with today observations.
The relic abundance of $\chi$ and $\bar\chi$ particles are basically determined by their annihilation cross section at the freeze-out epoch.
To investigate the effect of nonzero $\zeta$ in comparison to the case only with kinetic mixing, we compute the total $\chi \bar\chi$ annihilation cross section. The possible 2-body annihilation channels involve $f\bar{f}$, $W^+W^-$, $hh$, $ss$, $hs$, $Z^{(\prime)}Z^{(\prime)}$, $hZ^{(\prime)}$, and $sZ^{(\prime)}$. All these channels are mediated via $s$-channel $Z$ and $Z'$ bosons. In the case of $\zeta = 0$, all of these annihilation processes are controlled by a single parameter $t_{\varepsilon}$, such that they are typically suppressed by the observation that ${\sigma^{SI}_{\chi~N}}$ is very small. Here, we list two interaction vertices with larger contributions to the annihilation, 
\begin{equation}    
\mathcal{L} \supset  g_{Z^\prime W^+ W^-} \partial_\mu Z^{\prime\mu} W^+ W^- + g_{Z^\prime Z h} h Z_\mu  Z^{\prime\mu},
\end{equation}
where
\begin{eqnarray} 
g_{Z^\prime W^+ W^-}  &=&  \frac{{\hat c}_\mathrm{W} s_\xi e}{{\hat s}_\mathrm{W}},
\\
g_{Z^\prime Z h} &=&  -{\widetilde g}_0 c_\eta {\tilde c}^+_\xi {\tilde s}^-_\xi 
+  g^2_X  v_S \frac{ s_\eta s_{2\xi}}{ c_\epsilon^2}  
 - \zeta g_X \frac{ e v c_\eta ( c_\xi {\tilde c}^+_\xi -  s_\xi {\tilde s}^-_\xi ) }{ { \hat s}_\mathrm{W} {\hat c}_\mathrm{W} c_\epsilon} 
+ \zeta^2 g^2_X \frac{ v c_\eta s_{2\xi}}{ c^2_\epsilon},
\end{eqnarray}
with ${\widetilde g}_0 = e^2 v /( 2 {\hat s}^2_W {\hat c}^2_W )$. One could notice the presence of the extra parameter $\zeta$, which would mitigate the tension between direct detection and relic abundance. This can be confirmed by the relic density plotted with adjusted $\zeta$ in Fig.~\ref{relicenhanced}.

The calculation of the DM relic abundance in our model resorts to numerical procedures, where \texttt{micrOmegas}~\cite{micromegas2,micromegas3} is invoked, and Eq.~\eqref{xschiN} is coded into this framework after some double-checks.  
The attempts to globally explore all the allowed parameter regions is still restrained, for the numerical scans exhaust too much, especially when there are too many free parameters. In order to highlight the effect of nonzero $\zeta$, the results in Ref.~\cite{ZHHH:2020dm} for $\zeta = 0$ can be taken as a typical reference.
To this end, we prepare a scan over the model parameters, each round of the scan starts from a sampling of parameters $\{g_X, s_\epsilon, s_\xi\}$ running from small to large\footnote{$s_\xi$ starts from a value satisfying $\zeta=0$ in Eq.~\eqref{zetafromXi}.}.
We just fix $M_{Z^\prime} = 500~\si{GeV}$, and $\chi\chi$ annihilation would meet the $Z'$ resonance for $m_\chi \sim m_{Z'}/2$.

Given a point in this 3D space, the LEP and LHC constraints mentioned above are calculated at first (a failing parameter will be rejected hereafter), and then the effective DM-nucleon cross section is calculated and required to satisfy the LZ constraint~\cite{LUX-ZEPLIN:2022xrq}. Finally, a survival point will be fed to the estimation of the relic abundance $\Omega h^2$ comparing with the observed value $\Omega_0 h^2=0.1200\pm0.0012$~\cite{Aghanim:2018eyx}.

The scan started from $m_\chi = ( m_h + m_Z )/2 \simeq 115~\si{GeV}$ but the relic abundance is not satisfied for such low $m_\chi$. Up to $m_\chi = 210~\si{GeV}$, as shown in Fig.~\ref{relicenhanced}, the relic abundance $\Omega h^2$ has almost (but not yet, $\log(\Omega/\Omega_0 ) \sim 0.4$) reached $0.12$. Nonetheless, such a figure demonstrates that, for $g_X = 0.04$, $0.08$, $0.126$, and $0.4$, the obtained relic density for nonzero $\zeta$ (blue points) can be decreased by at least two orders of magnitude, comparing to the black points for $\zeta=0$.

The first physical solution, which passes all the constraints mentioned above and satisfies $|\Omega h^2 - 0.12| \le 0.012$, is found until $m_\chi = 215~\si{GeV}$, as shown in Tab.~\ref{relicSurvival}, 
When the DM candidate become heavier than $260~\si{GeV}$, which is the last row in this table, physical solutions will disappear again. In between, e.g., $m_\chi \sim 235~ \si{GeV}$, there are too many solutions to be recorded with $g_X$ running from ${10}^{-1}$ to ${10}^{-3}$. We think it just reflects the the $Z'$ resonance effect ( i.e. $2 m_chi \simeq m_{Z^\prime}$ )~\cite{resonacexception,Feng:2017drg} for freeze-out DM. In the case of $\zeta = 0$, the resonance region is around $m_\chi \sim 230\text{--}250~\si{GeV}$, as shown in Fig.~4 of Ref.~\cite{ZHHH:2020dm}.
The consequence of nonzero $\zeta$ is to extend the window to $215\text{--}260~\si{GeV}$. 

\begin{table}
\begin{center}
\setlength\tabcolsep{.6em}
\renewcommand{\arraystretch}{1.3}
\caption{The parameters corresponding to the correct relic density and consistent with the constraints. The unshown parameters take values from Ref.~\cite{ZHHH:2020dm}. }
\label{relicSurvival}
\begin{tabular}{  c  c  c  c }
\hline
 $m_\chi$ & $g_X$ & $s_\varepsilon$  & $\zeta$      \\
\hline
 $215$  & $6.31 \times{10}^{-1}$ & $5.13 \times{10}^{-2}$ &  $5.49 \times{10}^{-2}$~  \\ 
 $220$ & $3.98 \times{10}^{-1}$  & $-6.03 \times{10}^{-2}$ & $-5.96 \times{10}^{-2}$   \\
 $250$ & $5.01 \times{10}^{-3}$ & $-1.03 \times{10}^{-3}$ & $2.70 \times{10}^{-1}$  \\
 $260$ & $2.512$          & $-5.25 \times{10}^{-3}$ & $-1.42 \times{10}^{-3}$  \\
 \hline
\end{tabular}
\end{center}
\end{table} 

\begin{table}
\begin{center}
\setlength\tabcolsep{.6em}
\renewcommand{\arraystretch}{1.3}
\caption{The parameters corresponding to the correct relic density and consistent with the constraints for $m_\chi = ( m_h + m_Z )/2 \simeq 115~\si{GeV}$ and $m_{Z^\prime}=2.05 m_\chi$. }
\label{relicbosons}
\begin{tabular}{  c  c  c }
\hline
 $g_X$ & $s_\varepsilon$  & $\zeta$      \\
\hline
 $1.0 \times{10}^{-1}$~ & ~$-1.54\times{10}^{-3}$~ &  ~$-4.33 \times{10}^{-4}$~  \\ 
 $1.0 \times{10}^{-2}$  & $-1.83 \times{10}^{-3}$ & ~$-1.03\times{10}^{-1}$   \\
 $2.51 \times{10}^{-3}$ & $-1.0 \times{10}^{-3}$ & $-4.15 \times{10}^{-1}$  \\
  \hline
\end{tabular}
\end{center}
\end{table}

We also rescan around $m_\chi \simeq 115$ GeV but with $m_{Z^\prime}=2.05 m_\chi$ and just show a few from many solutions in Tab.~\ref{relicbosons}.
The relation $m_{Z^\prime}=2.05 m_\chi$ makes sure that the $Z'$ resonance effect is always important at the freeze-out epoch. In the $\zeta = 0$ case, this solution is totally rejected by the direct detection constraints, as shown in Fig.~5(a) of Ref.~\cite{ZHHH:2020dm}.
But in this work, it is recovered with a tuning of $\zeta$.

With \texttt{micrOmegas}, the $\gamma$-ray spectrum for DM indirect detection is also investigated for comparing with the upper bounds from Fermi-LAT $\gamma$-ray observations~\cite{Ackermann:2015zua}.
At least upon the parameters passing the above procedures in Tabs.~\ref{relicSurvival} and \ref{relicbosons}, there is no excess observed.

\section{Conclusions}
\label{sec:concl}

In this work, we introduce an extra $\mathrm{U}(1)_\mathrm{X}$ gauge symmetry, which is responsible for the interactions of Dirac fermionic DM with a $\mathrm{U}(1)_\mathrm{X}$ charge.
In particular, we assume the SM Higgs doublet also carry a $\mathrm{U}(1)_\mathrm{X}$ charge $\zeta$.
In order to make the SM Yukawa interaction terms gauge-invariant and make the theory free from gauge anomalies, the SM fermions are assigned $Y$-sequential $\mathrm{U}(1)_\mathrm{X}$ charges proportional to $\zeta$.
The mixing between the $\mathrm{U}(1)_\mathrm{X}$ and $\mathrm{U}(1)_\mathrm{Y}$ gauge fields are induced by both kinetic mixing and the $\mathrm{U}(1)_\mathrm{X}$ charge of the SM Higgs doublet.
Thus, the DM interactions with SM particles mediated by the $Z$ boson and the new $Z'$ boson are essentially controlled by the kinetic mixing parameter $\epsilon$ and the Higgs charge $\zeta$.

After the analytical calculation, we have performed numerical scans with fixed $m_{Z^\prime}$ over the parameter space of $g_X$, $s_\epsilon$, and $\zeta$. The new parameter $\zeta$ is found to invite  destructive interference in the effective DM-nucleon cross section, and it can affect the relic density by about two orders of magnitude. Although the magnitude of $\zeta$ is quite small due to the constraints from LEP and LHC experimental data, the cancellation between the interactions originated from the kinetic mixing and the Higgs $\mathrm{U}(1)_\mathrm{X}$ charge indeed takes place, 
Therefore, it can definitely extend the physical windows where the resonance effect for relic density are important. Nonetheless, the introduction of $\zeta$ itself is not enough to make generic parameter regions work.

\begin{acknowledgments}

The author L.Y. SHAN would thank Prof. Ying ZHANG for discussions about general $\mathrm{U}(1)_\mathrm{X}$ from the viewpoint of Effective Field Theory (Stueckelberg mechanism). The work is supported by National Natural Science Foundation of China under the grant NSFC-12135014.

\end{acknowledgments}

\appendix

\section{Parameter relations}
\label{param_relation}

By defining $\hat{m}_{Z}^{2} \equiv (\hat{g}^{2}+\hat{g}^{\prime 2})v^{2}/4$ , $\hat{m}_{Z^{\prime}}^{2} \equiv g_{X}^{2} ( v_{S}^{2} + \zeta^2 v^2)$, the masses squared of the $Z$ and $Z'$ bosons can be expressed as
\begin{equation}
m^2_{Z} = \hat{m}^2_{Z}  [ 1 + C_{Z}( s_{\epsilon}, \zeta) ~ ]  ,  ~~
m^2_{Z^{\prime}} = \frac{ \hat{m}^2_{Z^{\prime} } }{  [ 1 + C_{Z^{\prime}}( s_{\epsilon}, \zeta) ~ ] c^2_\epsilon },
\label{zsmass}
\end{equation}
where the small corrections are recast into
\begin{equation} \label{cZZp} 
C_Z  = {\cal Z} \hat{s}_\mathrm{W} t_{\xi}  ,\quad
~~ {\cal Z} = t_\epsilon - 2 \zeta \frac{ g_X \hat{c}_\mathrm{W} }{\sqrt{4\pi \alpha} c_\epsilon } ,\quad
C_{Z^\prime} =  C_Z + \frac{ 4 \zeta^2 t_\xi \hat{s}^2_\mathrm{W} g^2_X }{ \hat{g}^{\prime 2} [ t_\xi - {\cal Z} \hat{s}_\mathrm{W} ] c^2_\epsilon }
\end{equation}
It is straightforward to crosscheck that $C_{Z,Z^\prime}$ approaches to $ \hat{s}_W t_\epsilon t_\xi$ as Eq.~(20) of Ref.~\cite{ZHHH:2020dm} at the limit $\zeta\rightarrow 0 $.
From  Eqs.~(\ref{defXi}), (\ref{weak_mixing_angle}), and Eq.(\ref{cZZp}) the presence of $\zeta$ is seen to make nested functions more tanglesome, and to obstruct a direct eliminating of $C_Z$ or ${\hat s}_\mathrm{W}$. Below, an iterative approach is adopted to solve these nested functions. 

Firstly, $t_{2\xi}$ can be shaped as a Taylor expansion around $(C_Z,{\hat s}_\mathrm{W} ) \simeq ( 0, s_\mathrm{W} ) $ : 
\begin{eqnarray}
t^{(n+1)}_{2\xi} &=& \frac{2 {\cal Z} s_\mathrm{W} }{ 1- r } - \frac{ 2 (1+r) {\cal Z} s_\mathrm{W} }{{ (  1 - r )}^2 } C^{(n)}_Z + 
( \hat{s}^{(n)}_\mathrm{W} - s_\mathrm{W} )  \frac{ \partial t_{2\xi} }{\partial \hat{s}_\mathrm{W} }{\bigg\vert}^{\hat{s}_\mathrm{W} = s_\mathrm{W}}_{C_Z=0}  \nonumber  \\  
&& + {\cal Z}\cdot {\cal{O} } \big(  C^2_Z,~ ~ {( \hat{s}_\mathrm{W} - s_\mathrm{W})}^2,~ ( \hat{s}_\mathrm{W} - s_\mathrm{W}) C_Z \big).
\label{xiiter}
\end{eqnarray}
where $(n+1)$ denoting an approximation after the $n^\text{th}$ iteration. As confirmed below, it means that the contributions from higher orders will be incorporated by balancing the expression  with $n$ increasing. The expansion is based on the fact that $C_Z$ is constrained to be small because the $Z$ boson mass $m_Z$ has been well-measured, as well as the fact that $\hat{s}_\mathrm{W}$ is known to be very close to $s_\mathrm{W}$. More expressions are need to keep the iteration system close:
\begin{eqnarray}
 C^{(n+1)}_Z &=&  {\cal Z} s_\mathrm{W} t^{(n)}_{\xi} 
 + ( \hat{s}^{(n)}_\mathrm{W} - s_\mathrm{W}) t^{(n)}_{\xi} \frac{\partial ( {\cal Z} \hat{s}_\mathrm{W}) }{\partial \hat{s}_\mathrm{W} }{\bigg\vert}_{\hat{s}_\mathrm{W} = s_\mathrm{W}} 
+  {\cal{O} } \big( t^2_\xi, ~ {( \hat{s}_\mathrm{W} - s_\mathrm{W})}^2 \big), 
\label{Cziter}
\\
C^{(n+1)}_{Z^\prime} &=& C^{(n+1)}_Z + \frac{ \zeta^2 {\bar c}^2_\epsilon s^2_\mathrm{W} c^2_\mathrm{W} g^2_X t^{(n)}_{\xi} }{ 4 \pi\alpha [ t^{(n)}_{\xi} - {\cal Z} s_\mathrm{W} ] } + \zeta^2 \cdot {\cal{O} } \left( t^2_\xi, ~( \hat{s}_\mathrm{W} - s_\mathrm{W}) \right).
\label{Czpiter}
\end{eqnarray}

The leading iteration simply starts from
\begin{equation}
\label{leadingXi}
t^{(1)}_{\xi} = \frac{ {\cal Z} s_\mathrm{W} }{ 1- r }.
\end{equation}
This is also the leading expression in many previous studies. Together with $\hat{s}^{(1)}_\mathrm{W} = s_\mathrm{W}$, it can be utilized to obtain
\begin{equation} 
\label{leadingCz}
C^{(2)}_Z = \frac{ {\cal Z}^2 s^2_\mathrm{W} }{ 1- r }, ~~ 
{{\hat s}_\mathrm{W}}^{2~(2)} = s^2_\mathrm{W} +  \frac{ {\cal Z}^2 s^3_\mathrm{W} c^2_\mathrm{W} }{ ( 1- r  ) ( c^2_\mathrm{W} - s^2_\mathrm{W} ) }.
\end{equation}
When Eqs.~(\ref{leadingCz}) and (\ref{leadingXi}) are inserted back into Eq.~(\ref{xiiter}), one can get $t^{(3)}_{2\xi}$ in Eq.~(\ref{prxmtXi}). This expression can be expanded longer and longer when the round of iterations is further extended. These formulas are helpful to demonstrate the effects of a nonzero $\zeta$.

Alternatively, since the rotation angle $\xi$ takes more places in the Lagrangian, for example in the interactions among SM particles, especially in the fermion sector, while $\zeta$ only explicitly shows up in the $Z^\prime$ interactions in the Higgs sector, it is convenient to choose $\xi$ as a free parameter and regard $\zeta$ as a derived parameter. Along this line, Eqs.~(\ref{defXi}) and (\ref{cZZp}) are helpful in eliminating $\zeta$ (and even ${\hat s}_\mathrm{W}$): 
\begin{equation}
 C_Z = \frac{ t_\xi t_{2\xi} ( 1 - r )}{ 2 + ( 1+ r) t_\xi }.
\label{beuty}
\end{equation}
In such a choice one needs no more than the renormalization of the Weinberg angle ${\hat \theta}_\mathrm{W}$ via Eq.~(\ref{weak_mixing_angle}) with a small correction from $C_Z$:
\begin{eqnarray}
\hat{s}^2_\mathrm{W} &=& s^2_\mathrm{W} + \frac{1}{2}[ ( c^2_\mathrm{W} - s^2_\mathrm{W} ) \pm \sqrt{ { (c^2_\mathrm{W} - s^2_\mathrm{W} )}^2 - 4 s^2_\mathrm{W} c^2_\mathrm{W} C_Z } ] \\  \nonumber 
&=& s^2_\mathrm{W} + \frac{s^2_\mathrm{W} c^2_\mathrm{W} C_Z }{ ( c^2_\mathrm{W} - s^2_\mathrm{W} ) } - \frac{s^4_\mathrm{W} c^4_\mathrm{W} C^2_Z }{ { ( c^2_\mathrm{W} - s^2_\mathrm{W} ) }^3} + {\cal{O} } ( C^3_Z ).
\end{eqnarray}
In the case where $\zeta$ is explicitly involved, it can be recalculated from  
\begin{equation}
\label{zetafromXi}
\zeta = \frac{\sqrt{4\pi \alpha} c_\epsilon }{ 2 g_X \hat{c}_\mathrm{W}} \left\{ t_\epsilon - \frac{t_{2\xi} ( 1 - r )}{ {\hat s}_\mathrm{W}  [ 2 + ( 1+ r) t_\xi ]}   \right\} .
%\frac{ \sqrt{4\pi \alpha} c_\epsilon (t_\epsilon t_\xi {\hat s}_W - C_Z ) }{ 2 C_Z g_X {\hat c}_W {\hat s}_W t_\xi   } 
\end{equation}
Therein, ${\hat s}_\mathrm{W}$ (${\hat c}_\mathrm{W}$) should be replaced with the above formulation. It is possible that $\zeta$ could be not small when $r$ becomes large.

It is also straightforward to iterate $C_{Z^\prime}$ via Eq.~(\ref{Czpiter}), solve
 \begin{equation}\label{v_S}
 v^2_{S}=\frac{ m^2_{Z^{\prime}} c^2_\epsilon}{g^2_{X}}\left[ 1 + \left( C_{Z^\prime} - \frac{\zeta^2 g^2_X }{ c^2_\epsilon }\frac{v^2}{m^2_{Z^\prime}} \right) \right],
 \end{equation}
and propagate the basic parameters and corrections to the Higgs sector via
 \begin{eqnarray}
\lambda_H &=&\frac{(m_s^2+m_h^2)-c_{2\eta}(m_s^2-m_h^2)}{2v^2},\\
\lambda_S &=&\frac{(m_s^2+m_h^2)+c_{2\eta}(m_s^2-m_h^2)}{2v^2},\\
\lambda_{HS} &=&\frac{t_{2\eta}(\lambda_Hv^2-\lambda_Sv_S^2)}{2vv_S}.
\end{eqnarray}

Utilizing the above relations, all the parameters in the model are calculable on the base of $g_X$, $m_{Z^\prime}$, $s_{\epsilon}$, $\zeta$, $m_s$, and $s_{\eta}$, together with the well-measured parameters $G_\mathrm{F}$, $m_Z$, and $\alpha$. 
 
\bibliographystyle{utphys}
\bibliography{reference}

\providecommand{\href}[2]{#2}\begingroup\raggedright\begin{thebibliography}{10}

\bibitem{Jungman:1995df}
G.~Jungman, M.~Kamionkowski, and K.~Griest, ``{Supersymmetric dark matter},''
  \href{http://dx.doi.org/10.1016/0370-1573(95)00058-5}{{\em Phys. Rept.}
  {\bfseries 267} (1996) 195--373},
  \href{http://arxiv.org/abs/hep-ph/9506380}{{\ttfamily arXiv:hep-ph/9506380}}.

\bibitem{Bertone:2004pz}
G.~Bertone, D.~Hooper, and J.~Silk, ``{Particle dark matter: Evidence,
  candidates and constraints},''
  \href{http://dx.doi.org/10.1016/j.physrep.2004.08.031}{{\em Phys. Rept.}
  {\bfseries 405} (2005) 279--390},
  \href{http://arxiv.org/abs/hep-ph/0404175}{{\ttfamily arXiv:hep-ph/0404175}}.

\bibitem{Feng:2010gw}
J.~L. Feng, ``{Dark Matter Candidates from Particle Physics and Methods of
  Detection},''
  \href{http://dx.doi.org/10.1146/annurev-astro-082708-101659}{{\em Ann. Rev.
  Astron. Astrophys.} {\bfseries 48} (2010) 495--545},
  \href{http://arxiv.org/abs/1003.0904}{{\ttfamily arXiv:1003.0904
  [astro-ph.CO]}}.

\bibitem{Young:2016ala}
B.-L. Young, ``{A survey of dark matter and related topics in cosmology},''
  \href{http://dx.doi.org/10.1007/s11467-016-0583-4}{{\em Front. Phys.
  (Beijing)} {\bfseries 12} (2017) 121201}. [Erratum: Front.Phys.(Beijing) 12,
  121202 (2017)].

\bibitem{Langacker:2008yv}
P.~Langacker, ``{The Physics of Heavy $Z^\prime$ Gauge Bosons},''
  \href{http://dx.doi.org/10.1103/RevModPhys.81.1199}{{\em Rev. Mod. Phys.}
  {\bfseries 81} (2009) 1199--1228},
  \href{http://arxiv.org/abs/0801.1345}{{\ttfamily arXiv:0801.1345 [hep-ph]}}.

\bibitem{Feldman:2007wj}
D.~Feldman, Z.~Liu, and P.~Nath, ``{The Stueckelberg Z-prime Extension with
  Kinetic Mixing and Milli-Charged Dark Matter From the Hidden Sector},''
  \href{http://dx.doi.org/10.1103/PhysRevD.75.115001}{{\em Phys. Rev. D}
  {\bfseries 75} (2007) 115001},
  \href{http://arxiv.org/abs/hep-ph/0702123}{{\ttfamily arXiv:hep-ph/0702123}}.

\bibitem{Pospelov:2007mp}
M.~Pospelov, A.~Ritz, and M.~B. Voloshin, ``{Secluded WIMP Dark Matter},''
  \href{http://dx.doi.org/10.1016/j.physletb.2008.02.052}{{\em Phys. Lett. B}
  {\bfseries 662} (2008) 53--61},
  \href{http://arxiv.org/abs/0711.4866}{{\ttfamily arXiv:0711.4866 [hep-ph]}}.

\bibitem{Mambrini:2010dq}
Y.~Mambrini, ``{The Kinetic dark-mixing in the light of CoGENT and XENON100},''
  \href{http://dx.doi.org/10.1088/1475-7516/2010/09/022}{{\em JCAP} {\bfseries
  09} (2010) 022}, \href{http://arxiv.org/abs/1006.3318}{{\ttfamily
  arXiv:1006.3318 [hep-ph]}}.

\bibitem{Kang:2010mh}
Z.~Kang, T.~Li, T.~Liu, C.~Tong, and J.~M. Yang, ``{Light Dark Matter from the
  $U(1)_X$ Sector in the NMSSM with Gauge Mediation},''
  \href{http://dx.doi.org/10.1088/1475-7516/2011/01/028}{{\em JCAP} {\bfseries
  01} (2011) 028}, \href{http://arxiv.org/abs/1008.5243}{{\ttfamily
  arXiv:1008.5243 [hep-ph]}}.

\bibitem{Chun:2010ve}
E.~J. Chun, J.-C. Park, and S.~Scopel, ``{Dark matter and a new gauge boson
  through kinetic mixing},''
  \href{http://dx.doi.org/10.1007/JHEP02(2011)100}{{\em JHEP} {\bfseries 02}
  (2011) 100}, \href{http://arxiv.org/abs/1011.3300}{{\ttfamily arXiv:1011.3300
  [hep-ph]}}.

\bibitem{Mambrini:2011dw}
Y.~Mambrini, ``{The ZZ' kinetic mixing in the light of the recent direct and
  indirect dark matter searches},''
  \href{http://dx.doi.org/10.1088/1475-7516/2011/07/009}{{\em JCAP} {\bfseries
  07} (2011) 009}, \href{http://arxiv.org/abs/1104.4799}{{\ttfamily
  arXiv:1104.4799 [hep-ph]}}.

\bibitem{Frandsen:2011cg}
M.~T. Frandsen, F.~Kahlhoefer, S.~Sarkar, and K.~Schmidt-Hoberg, ``{Direct
  detection of dark matter in models with a light Z'},''
  \href{http://dx.doi.org/10.1007/JHEP09(2011)128}{{\em JHEP} {\bfseries 09}
  (2011) 128}, \href{http://arxiv.org/abs/1107.2118}{{\ttfamily arXiv:1107.2118
  [hep-ph]}}.

\bibitem{Gao:2011ka}
X.~Gao, Z.~Kang, and T.~Li, ``{Origins of the Isospin Violation of Dark Matter
  Interactions},'' \href{http://dx.doi.org/10.1088/1475-7516/2013/01/021}{{\em
  JCAP} {\bfseries 01} (2013) 021},
  \href{http://arxiv.org/abs/1107.3529}{{\ttfamily arXiv:1107.3529 [hep-ph]}}.

\bibitem{Chu:2011be}
X.~Chu, T.~Hambye, and M.~H.~G. Tytgat, ``{The Four Basic Ways of Creating Dark
  Matter Through a Portal},''
  \href{http://dx.doi.org/10.1088/1475-7516/2012/05/034}{{\em JCAP} {\bfseries
  05} (2012) 034}, \href{http://arxiv.org/abs/1112.0493}{{\ttfamily
  arXiv:1112.0493 [hep-ph]}}.

\bibitem{Frandsen:2012rk}
M.~T. Frandsen, F.~Kahlhoefer, A.~Preston, S.~Sarkar, and K.~Schmidt-Hoberg,
  ``{LHC and Tevatron Bounds on the Dark Matter Direct Detection Cross-Section
  for Vector Mediators},''
  \href{http://dx.doi.org/10.1007/JHEP07(2012)123}{{\em JHEP} {\bfseries 07}
  (2012) 123}, \href{http://arxiv.org/abs/1204.3839}{{\ttfamily arXiv:1204.3839
  [hep-ph]}}.

\bibitem{Jia:2013lza}
L.-B. Jia and X.-Q. Li, ``{Study of a WIMP dark matter model with the updated
  results of CDMS II},''
  \href{http://dx.doi.org/10.1103/PhysRevD.89.035006}{{\em Phys. Rev. D}
  {\bfseries 89} (2014) 035006},
  \href{http://arxiv.org/abs/1309.6029}{{\ttfamily arXiv:1309.6029 [hep-ph]}}.

\bibitem{Belanger:2013tla}
G.~B\'elanger, A.~Goudelis, J.-C. Park, and A.~Pukhov, ``{Isospin-violating
  dark matter from a double portal},''
  \href{http://dx.doi.org/10.1088/1475-7516/2014/02/020}{{\em JCAP} {\bfseries
  02} (2014) 020}, \href{http://arxiv.org/abs/1311.0022}{{\ttfamily
  arXiv:1311.0022 [hep-ph]}}.

\bibitem{Chen:2014tka}
N.~Chen, Q.~Wang, W.~Zhao, S.-T. Lin, Q.~Yue, and J.~Li, ``{Exothermic
  isospin-violating dark matter after SuperCDMS and CDEX},''
  \href{http://dx.doi.org/10.1016/j.physletb.2015.02.043}{{\em Phys. Lett. B}
  {\bfseries 743} (2015) 205--212},
  \href{http://arxiv.org/abs/1404.6043}{{\ttfamily arXiv:1404.6043 [hep-ph]}}.

\bibitem{Arcadi:2017kky}
G.~Arcadi, M.~Dutra, P.~Ghosh, M.~Lindner, Y.~Mambrini, M.~Pierre, S.~Profumo,
  and F.~S. Queiroz, ``{The waning of the WIMP? A review of models, searches,
  and constraints},''
  \href{http://dx.doi.org/10.1140/epjc/s10052-018-5662-y}{{\em Eur. Phys. J. C}
  {\bfseries 78} (2018) 203}, \href{http://arxiv.org/abs/1703.07364}{{\ttfamily
  arXiv:1703.07364 [hep-ph]}}.

\bibitem{Liu:2017lpo}
J.~Liu, X.-P. Wang, and F.~Yu, ``{A Tale of Two Portals: Testing Light, Hidden
  New Physics at Future $e^+ e^-$ Colliders},''
  \href{http://dx.doi.org/10.1007/JHEP06(2017)077}{{\em JHEP} {\bfseries 06}
  (2017) 077}, \href{http://arxiv.org/abs/1704.00730}{{\ttfamily
  arXiv:1704.00730 [hep-ph]}}.

\bibitem{Dutra:2018gmv}
M.~Dutra, M.~Lindner, S.~Profumo, F.~S. Queiroz, W.~Rodejohann, and
  C.~Siqueira, ``{MeV Dark Matter Complementarity and the Dark Photon
  Portal},'' \href{http://dx.doi.org/10.1088/1475-7516/2018/03/037}{{\em JCAP}
  {\bfseries 03} (2018) 037}, \href{http://arxiv.org/abs/1801.05447}{{\ttfamily
  arXiv:1801.05447 [hep-ph]}}.

\bibitem{Bauer:2018egk}
M.~Bauer, S.~Diefenbacher, T.~Plehn, M.~Russell, and D.~A. Camargo, ``{Dark
  Matter in Anomaly-Free Gauge Extensions},''
  \href{http://dx.doi.org/10.21468/SciPostPhys.5.4.036}{{\em SciPost Phys.}
  {\bfseries 5} (2018) 036}, \href{http://arxiv.org/abs/1805.01904}{{\ttfamily
  arXiv:1805.01904 [hep-ph]}}.

\bibitem{Koren:2019iuv}
S.~Koren and R.~McGehee, ``{Freezing-in twin dark matter},''
  \href{http://dx.doi.org/10.1103/PhysRevD.101.055024}{{\em Phys. Rev. D}
  {\bfseries 101} (2020) 055024},
  \href{http://arxiv.org/abs/1908.03559}{{\ttfamily arXiv:1908.03559
  [hep-ph]}}.

\bibitem{Jung:2020ukk}
D.-W. Jung, S.-H. Nam, C.~Yu, Y.~G. Kim, and K.~Y. Lee, ``{Singlet fermionic
  dark matter with dark $Z$},''
  \href{http://dx.doi.org/10.1140/epjc/s10052-020-8080-x}{{\em Eur. Phys. J. C}
  {\bfseries 80} (2020) 513}, \href{http://arxiv.org/abs/2002.10075}{{\ttfamily
  arXiv:2002.10075 [hep-ph]}}.

\bibitem{ZHHH:2020dm}
J.~Lao, C.~Cai, Z.-H. Yu, Y.-P. Zeng, and H.-H. Zhang, ``{Fermionic and scalar
  dark matter with hidden $\mathrm{U}(1)$ gauge interaction and kinetic
  mixing},'' \href{http://dx.doi.org/10.1103/PhysRevD.101.095031}{{\em Phys.
  Rev. D} {\bfseries 101} (2020) 095031},
  \href{http://arxiv.org/abs/2003.02516}{{\ttfamily arXiv:2003.02516
  [hep-ph]}}.

\bibitem{Cai:2021evx}
C.~Cai, Y.-P. Zeng, and H.-H. Zhang, ``{Cancellation mechanism of dark matter
  direct detection in Higgs-portal and vector-portal models},''
  \href{http://dx.doi.org/10.1007/JHEP01(2022)117}{{\em JHEP} {\bfseries 01}
  (2022) 117}, \href{http://arxiv.org/abs/2109.11499}{{\ttfamily
  arXiv:2109.11499 [hep-ph]}}.

\bibitem{Liu:2022evb}
D.-Y. Liu, C.~Cai, X.-M. Jiang, Z.-H. Yu, and H.-H. Zhang, ``{Ultraviolet
  completion of pseudo-Nambu-Goldstone dark matter with a hidden U(1) gauge
  symmetry},'' \href{http://dx.doi.org/10.1007/JHEP02(2023)104}{{\em JHEP}
  {\bfseries 02} (2023) 104}, \href{http://arxiv.org/abs/2208.06653}{{\ttfamily
  arXiv:2208.06653 [hep-ph]}}.

\bibitem{Carvunis:2022yur}
A.~Carvunis, N.~McGinnis, and D.~E. Morrissey, ``{Relic challenges for
  vector-like fermions as connectors to a dark sector},''
  \href{http://dx.doi.org/10.1007/JHEP01(2023)014}{{\em JHEP} {\bfseries 01}
  (2023) 014}, \href{http://arxiv.org/abs/2209.14305}{{\ttfamily
  arXiv:2209.14305 [hep-ph]}}.

\bibitem{Qiu:2023wbs}
Z.-Y. Qiu and Z.-H. Yu, ``{Gravitational waves from cosmic strings associated
  with pseudo-Nambu-Goldstone dark matter},''
  \href{http://dx.doi.org/10.1088/1674-1137/acd9bf}{{\em Chin. Phys. C}
  {\bfseries 47} (2023) 085104},
  \href{http://arxiv.org/abs/2304.02506}{{\ttfamily arXiv:2304.02506
  [hep-ph]}}.

\bibitem{Holdom:1985ag}
B.~Holdom, ``{Two U(1)'s and Epsilon Charge Shifts},''
  \href{http://dx.doi.org/10.1016/0370-2693(86)91377-8}{{\em Phys. Lett. B}
  {\bfseries 166} (1986) 196--198}.

\bibitem{YingZ:2009}
Y.~Zhang and Q.~Wang, ``{$Z'$ Boson Mixings with $Z$-$\gamma$ and charge
  assignments},'' \href{http://dx.doi.org/10.1088/1126-6708/2009/07/012}{{\em
  JHEP} {\bfseries 07} (2009) 012},
  \href{http://arxiv.org/abs/0904.2047}{{\ttfamily arXiv:0904.2047 [hep-ph]}}.

\bibitem{firstFO}
H.-Y. Chiu, ``{Symmetry between particle and anti-particle populations in the
  universe},'' \href{http://dx.doi.org/10.1103/PhysRevLett.17.712}{{\em Phys.
  Rev. Lett.} {\bfseries 17} (1966) 712}.

\bibitem{Kolb:1990vq}
E.~W. Kolb and M.~S. Turner, ``{The Early Universe},''
  \href{http://dx.doi.org/10.1201/9780429492860}{{\em Front. Phys.} {\bfseries
  69} (1990) 1--547}.

\bibitem{relicFO}
P.~Gondolo and G.~Gelmini, ``{Cosmic abundances of stable particles: Improved
  analysis},'' \href{http://dx.doi.org/10.1016/0550-3213(91)90438-4}{{\em Nucl.
  Phys. B} {\bfseries 360} (1991) 145--179}.

\bibitem{Ekstedt:2016}
A.~Ekstedt, R.~Enberg, G.~Ingelman, J.~L\"ofgren, and T.~Mandal,
  ``{Constraining minimal anomaly free $\mathrm{U}(1)$ extensions of the
  Standard Model},'' \href{http://dx.doi.org/10.1007/JHEP11(2016)071}{{\em
  JHEP} {\bfseries 11} (2016) 071},
  \href{http://arxiv.org/abs/1605.04855}{{\ttfamily arXiv:1605.04855
  [hep-ph]}}.

\bibitem{Fayet:1989mq}
P.~Fayet, ``{The Fifth Force Charge as a Linear Combination of Baryonic,
  Leptonic (Or $B - L $) and Electric Charges},''
  \href{http://dx.doi.org/10.1016/0370-2693(89)91294-X}{{\em Phys. Lett. B}
  {\bfseries 227} (1989) 127--132}.

\bibitem{Bogdan:2003}
T.~Appelquist, B.~A. Dobrescu, and A.~R. Hopper, ``{Nonexotic Neutral Gauge
  Bosons},'' \href{http://dx.doi.org/10.1103/PhysRevD.68.035012}{{\em Phys.
  Rev. D} {\bfseries 68} (2003) 035012},
  \href{http://arxiv.org/abs/hep-ph/0212073}{{\ttfamily arXiv:hep-ph/0212073}}.

\bibitem{Fairbairn:2017}
J.~Ellis, M.~Fairbairn, and P.~Tunney, ``{Anomaly-Free Dark Matter Models are
  not so Simple},'' \href{http://dx.doi.org/10.1007/JHEP08(2017)053}{{\em JHEP}
  {\bfseries 08} (2017) 053}, \href{http://arxiv.org/abs/1704.03850}{{\ttfamily
  arXiv:1704.03850 [hep-ph]}}.

\bibitem{Ellis:2018xal}
J.~Ellis, M.~Fairbairn, and P.~Tunney, ``{Phenomenological Constraints on
  Anomaly-Free Dark Matter Models},''
  \href{http://arxiv.org/abs/1807.02503}{{\ttfamily arXiv:1807.02503
  [hep-ph]}}.

\bibitem{Babu:1997st}
K.~S. Babu, C.~F. Kolda, and J.~March-Russell, ``{Implications of generalized Z
  - Z-prime mixing},'' \href{http://dx.doi.org/10.1103/PhysRevD.57.6788}{{\em
  Phys. Rev. D} {\bfseries 57} (1998) 6788--6792},
  \href{http://arxiv.org/abs/hep-ph/9710441}{{\ttfamily arXiv:hep-ph/9710441}}.

\bibitem{Carena:2004}
M.~Carena, A.~Daleo, B.~A. Dobrescu, and T.~M.~P. Tait, ``{$Z^\prime$ gauge
  bosons at the Tevatron},''
  \href{http://dx.doi.org/10.1103/PhysRevD.70.093009}{{\em Phys. Rev. D}
  {\bfseries 70} (2004) 093009},
  \href{http://arxiv.org/abs/hep-ph/0408098}{{\ttfamily arXiv:hep-ph/0408098}}.

\bibitem{adas:2016}
A.~Das, S.~Oda, N.~Okada, and D.-s. Takahashi, ``{Classically conformal U(1)'
  extended standard model, electroweak vacuum stability, and LHC Run-2
  bounds},'' \href{http://dx.doi.org/10.1103/PhysRevD.93.115038}{{\em Phys.
  Rev. D} {\bfseries 93} (2016) 115038},
  \href{http://arxiv.org/abs/1605.01157}{{\ttfamily arXiv:1605.01157
  [hep-ph]}}.

\bibitem{Das:2019pua}
A.~Das, S.~Goswami, K.~N. Vishnudath, and T.~Nomura, ``{Constraining a general
  U(1)$^\prime$ inverse seesaw model from vacuum stability, dark matter and
  collider},'' \href{http://dx.doi.org/10.1103/PhysRevD.101.055026}{{\em Phys.
  Rev. D} {\bfseries 101} (2020) 055026},
  \href{http://arxiv.org/abs/1905.00201}{{\ttfamily arXiv:1905.00201
  [hep-ph]}}.

\bibitem{Das:2022oyx}
A.~Das, S.~Gola, S.~Mandal, and N.~Sinha, ``{Two-component scalar and fermionic
  dark matter candidates in a generic U(1)X model},''
  \href{http://dx.doi.org/10.1016/j.physletb.2022.137117}{{\em Phys. Lett. B}
  {\bfseries 829} (2022) 137117},
  \href{http://arxiv.org/abs/2202.01443}{{\ttfamily arXiv:2202.01443
  [hep-ph]}}.

\bibitem{Fayet:1990wx}
P.~Fayet, ``{Extra U(1)'s and New Forces},''
  \href{http://dx.doi.org/10.1016/0550-3213(90)90381-M}{{\em Nucl. Phys. B}
  {\bfseries 347} (1990) 743--768}.

\bibitem{Fayet:2004bw}
P.~Fayet, ``{Light spin 1/2 or spin 0 dark matter particles},''
  \href{http://dx.doi.org/10.1103/PhysRevD.70.023514}{{\em Phys. Rev. D}
  {\bfseries 70} (2004) 023514}, \href{http://arxiv.org/abs/0403226
  [hep-ph]}{{\ttfamily arXiv:0403226 [hep-ph]}}.

\bibitem{Feng:2011vu}
J.~L. Feng, J.~Kumar, D.~Marfatia, and D.~Sanford, ``{Isospin-Violating Dark
  Matter},'' \href{http://dx.doi.org/10.1016/j.physletb.2011.07.083}{{\em Phys.
  Lett. B} {\bfseries 703} (2011) 124--127},
  \href{http://arxiv.org/abs/1102.4331}{{\ttfamily arXiv:1102.4331 [hep-ph]}}.

\bibitem{Higgs:1964ia}
P.~W. Higgs, ``{Broken symmetries, massless particles and gauge fields},''
  \href{http://dx.doi.org/10.1016/0031-9163(64)91136-9}{{\em Phys. Lett.}
  {\bfseries 12} (1964) 132--133}.

\bibitem{Higgs:1964pj}
P.~W. Higgs, ``{Broken Symmetries and the Masses of Gauge Bosons},''
  \href{http://dx.doi.org/10.1103/PhysRevLett.13.508}{{\em Phys. Rev. Lett.}
  {\bfseries 13} (1964) 508--509}.

\bibitem{Englert:1964et}
F.~Englert and R.~Brout, ``{Broken Symmetry and the Mass of Gauge Vector
  Mesons},'' \href{http://dx.doi.org/10.1103/PhysRevLett.13.321}{{\em Phys.
  Rev. Lett.} {\bfseries 13} (1964) 321--323}.

\bibitem{Stueckelberg:1938hvi}
E.~C.~G. Stueckelberg, ``{Interaction energy in electrodynamics and in the
  field theory of nuclear forces},''
  \href{http://dx.doi.org/10.5169/seals-110852}{{\em Helv. Phys. Acta}
  {\bfseries 11} (1938) 225--244}.

\bibitem{Chodos:1971yj}
A.~Chodos and F.~Cooper, ``{New lagrangian formalism for infinite-component
  field theories},'' \href{http://dx.doi.org/10.1103/PhysRevD.3.2461}{{\em
  Phys. Rev. D} {\bfseries 3} (1971) 2461--2472}.

\bibitem{weinbergAngle}
{\bfseries Particle Data Group} Collaboration, R.~L. Workman {\em et~al.},
  ``{Review of Particle Physics},''
  \href{http://dx.doi.org/10.1093/ptep/ptac097}{{\em PTEP} {\bfseries 2022}
  (2022) 083C01}.

\bibitem{Burgess:1993vc}
C.~P. Burgess, S.~Godfrey, H.~Konig, D.~London, and I.~Maksymyk, ``{Model
  independent global constraints on new physics},''
  \href{http://dx.doi.org/10.1103/PhysRevD.49.6115}{{\em Phys. Rev. D}
  {\bfseries 49} (1994) 6115--6147},
  \href{http://arxiv.org/abs/hep-ph/9312291}{{\ttfamily arXiv:hep-ph/9312291}}.

\bibitem{ATLAS:2019erb}
{\bfseries ATLAS} Collaboration, G.~Aad {\em et~al.}, ``{Search for high-mass
  dilepton resonances using 139 fb$^{-1}$ of $pp$ collision data collected at
  $\sqrt{s}=$13 TeV with the ATLAS detector},''
  \href{http://dx.doi.org/10.1016/j.physletb.2019.07.016}{{\em Phys. Lett. B}
  {\bfseries 796} (2019) 68--87},
  \href{http://arxiv.org/abs/1903.06248}{{\ttfamily arXiv:1903.06248
  [hep-ex]}}.

\bibitem{CMS:2021ctt}
{\bfseries CMS} Collaboration, A.~M. Sirunyan {\em et~al.}, ``{Search for
  resonant and nonresonant new phenomena in high-mass dilepton final states at
  $ \sqrt{s} $ = 13 TeV},''
  \href{http://dx.doi.org/10.1007/JHEP07(2021)208}{{\em JHEP} {\bfseries 07}
  (2021) 208}, \href{http://arxiv.org/abs/2103.02708}{{\ttfamily
  arXiv:2103.02708 [hep-ex]}}.

\bibitem{Zheng:2010js}
J.-M. Zheng, Z.-H. Yu, J.-W. Shao, X.-J. Bi, Z.~Li, and H.-H. Zhang,
  ``{Constraining the interaction strength between dark matter and visible
  matter: I. fermionic dark matter},''
  \href{http://dx.doi.org/10.1016/j.nuclphysb.2011.09.009}{{\em Nucl. Phys. B}
  {\bfseries 854} (2012) 350--374},
  \href{http://arxiv.org/abs/1012.2022}{{\ttfamily arXiv:1012.2022 [hep-ph]}}.

\bibitem{Feng:2013fyw}
J.~L. Feng, J.~Kumar, and D.~Sanford, ``{Xenophobic Dark Matter},''
  \href{http://dx.doi.org/10.1103/PhysRevD.88.015021}{{\em Phys. Rev. D}
  {\bfseries 88} (2013) 015021},
  \href{http://arxiv.org/abs/1306.2315}{{\ttfamily arXiv:1306.2315 [hep-ph]}}.

\bibitem{Backovic:2013dpa}
M.~Backovic, K.~Kong, and M.~McCaskey, ``{MadDM v.1.0: Computation of Dark
  Matter Relic Abundance Using MadGraph5},''
  \href{http://dx.doi.org/10.1016/j.dark.2014.04.001}{{\em Physics of the Dark
  Universe} {\bfseries 5-6} (2014) 18--28},
  \href{http://arxiv.org/abs/1308.4955}{{\ttfamily arXiv:1308.4955 [hep-ph]}}.

\bibitem{Ambrogi:2018jqj}
F.~Ambrogi, C.~Arina, M.~Backovic, J.~Heisig, F.~Maltoni, L.~Mantani,
  O.~Mattelaer, and G.~Mohlabeng, ``{MadDM v.3.0: a Comprehensive Tool for Dark
  Matter Studies},'' \href{http://dx.doi.org/10.1016/j.dark.2018.11.009}{{\em
  Phys. Dark Univ.} {\bfseries 24} (2019) 100249},
  \href{http://arxiv.org/abs/1804.00044}{{\ttfamily arXiv:1804.00044
  [hep-ph]}}.

\bibitem{PandaX-4T:2021bab}
{\bfseries PandaX-4T} Collaboration, Y.~Meng {\em et~al.}, ``{Dark Matter
  Search Results from the PandaX-4T Commissioning Run},''
  \href{http://dx.doi.org/10.1103/PhysRevLett.127.261802}{{\em Phys. Rev.
  Lett.} {\bfseries 127} (2021) 261802},
  \href{http://arxiv.org/abs/2107.13438}{{\ttfamily arXiv:2107.13438
  [hep-ex]}}.

\bibitem{LUX-ZEPLIN:2022xrq}
{\bfseries LUX-ZEPLIN} Collaboration, J.~Aalbers {\em et~al.}, ``{First Dark
  Matter Search Results from the LUX-ZEPLIN (LZ) Experiment},''
  \href{http://dx.doi.org/10.1103/PhysRevLett.131.041002}{{\em Phys. Rev.
  Lett.} {\bfseries 131} (2023) 041002},
  \href{http://arxiv.org/abs/2207.03764}{{\ttfamily arXiv:2207.03764
  [hep-ex]}}.

\bibitem{XENON:2023cxc}
{\bfseries XENON} Collaboration, E.~Aprile {\em et~al.}, ``{First Dark Matter
  Search with Nuclear Recoils from the XENONnT Experiment},''
  \href{http://dx.doi.org/10.1103/PhysRevLett.131.041003}{{\em Phys. Rev.
  Lett.} {\bfseries 131} (2023) 041003},
  \href{http://arxiv.org/abs/2303.14729}{{\ttfamily arXiv:2303.14729
  [hep-ex]}}.

\bibitem{micromegas2}
G.~Belanger, F.~Boudjema, A.~Pukhov, and A.~Semenov, ``{MicrOMEGAs 2.0: A
  Program to calculate the relic density of dark matter in a generic model},''
  \href{http://dx.doi.org/10.1016/j.cpc.2006.11.008}{{\em Comput. Phys.
  Commun.} {\bfseries 176} (2007) 367--382},
  \href{http://arxiv.org/abs/hep-ph/0607059}{{\ttfamily arXiv:hep-ph/0607059}}.

\bibitem{micromegas3}
G.~Belanger, F.~Boudjema, A.~Pukhov, and A.~Semenov, ``{micrOMEGAs$\_$3: A
  program for calculating dark matter observables},''
  \href{http://dx.doi.org/10.1016/j.cpc.2013.10.016}{{\em Comput. Phys.
  Commun.} {\bfseries 185} (2014) 960--985},
  \href{http://arxiv.org/abs/1305.0237}{{\ttfamily arXiv:1305.0237 [hep-ph]}}.

\bibitem{Aghanim:2018eyx}
{\bfseries Planck} Collaboration, N.~Aghanim {\em et~al.}, ``{Planck 2018
  results. VI. Cosmological parameters},''
  \href{http://dx.doi.org/10.1051/0004-6361/201833910}{{\em Astron. Astrophys.}
  {\bfseries 641} (2020) A6}, \href{http://arxiv.org/abs/1807.06209}{{\ttfamily
  arXiv:1807.06209 [astro-ph.CO]}}. [Erratum: Astron.Astrophys. 652, C4
  (2021)].

\bibitem{resonacexception}
K.~Griest and D.~Seckel, ``{Three exceptions in the calculation of relic
  abundances},'' \href{http://dx.doi.org/10.1103/PhysRevD.43.3191}{{\em Phys.
  Rev. D} {\bfseries 43} (1991) 3191--3203}.

\bibitem{Feng:2017drg}
J.~L. Feng and J.~Smolinsky, ``{Impact of a resonance on thermal targets for
  invisible dark photon searches},''
  \href{http://dx.doi.org/10.1103/PhysRevD.96.095022}{{\em Phys. Rev. D}
  {\bfseries 96} (2017) 095022},
  \href{http://arxiv.org/abs/1707.03835}{{\ttfamily arXiv:1707.03835
  [hep-ph]}}.

\bibitem{Ackermann:2015zua}
{\bfseries Fermi-LAT} Collaboration, M.~Ackermann {\em et~al.}, ``{Searching
  for Dark Matter Annihilation from Milky Way Dwarf Spheroidal Galaxies with
  Six Years of Fermi Large Area Telescope Data},''
  \href{http://dx.doi.org/10.1103/PhysRevLett.115.231301}{{\em Phys. Rev.
  Lett.} {\bfseries 115} (2015) 231301},
  \href{http://arxiv.org/abs/1503.02641}{{\ttfamily arXiv:1503.02641
  [astro-ph.HE]}}.

\end{thebibliography}\endgroup

\end{document}